\documentclass[letter]{aa} 
\usepackage{graphicx}
\usepackage{url}
\usepackage{subcaption}
\usepackage[noindentafter]{titlesec}
\usepackage{enumitem}

\usepackage{txfonts}
\usepackage[breaklinks,colorlinks,citecolor=blue,pdfa=true,backref=page]{hyperref}

\usepackage{natbib,twoopt}
\usepackage{longtable}
\newcommand{\GG}[1]{}

\graphicspath{{images/}}

\bibpunct{(}{)}{;}{a}{}{,} 
\makeatletter
\newcommandtwoopt{\citeads}[3][][]{\href{http://adsabs.harvard.edu/abs/#3}%
{\def\hyper@linkstart##1##2{}%
\let\hyper@linkend\@empty\citealp[#1][#2]{#3}}}
\newcommandtwoopt{\citepads}[3][][]{\href{http://adsabs.harvard.edu/abs/#3}%
{\def\hyper@linkstart##1##2{}%
\let\hyper@linkend\@empty\citep[#1][#2]{#3}}}
\newcommandtwoopt{\citetads}[3][][]{\href{http://adsabs.harvard.edu/abs/#3}%
{\def\hyper@linkstart##1##2{}%
\let\hyper@linkend\@empty\citet[#1][#2]{#3}}}
\newcommandtwoopt{\citeyearads}[3][][]%
{\href{http://adsabs.harvard.edu/abs/#3}
{\def\hyper@linkstart##1##2{}%
\let\hyper@linkend\@empty\citeyear[#1][#2]{#3}}}
\makeatother

\def \arcsec      {\text{$^{\prime\prime}$}}

\begin{document}

\title{Search and analysis of giant radio galaxies with associated nuclei (SAGAN)}

\subtitle{IV. Interplay with the Supercluster environment}

   \titlerunning{SAGAN-IV: Interplay with the supercluster environment}

\author {Shishir Sankhyayan\inst{1}\thanks{\email{\url{shishir.sankhyayan@ut.ee}}}
\and Pratik Dabhade\inst{2,3}\thanks{\email{\url{pratik.dabhade@iac.es}}}
}
  
\authorrunning{S. Sankhyayan and P. Dabhade}
\institute{$^{1}$Tartu Observatory, University of Tartu, Observatooriumi~1, 61602 T\~oravere, Estonia  \\
$^{2}$Instituto de Astrof\' isica de Canarias, Calle V\' ia L\'actea, s/n, E-38205, La Laguna, Tenerife, Spain\\
$^{3}$Universidad de La Laguna (ULL), Departamento de Astrofisica, La Laguna,
E-38206, Tenerife, Spain\\ 
}

 \date{\today} 
 
 \abstract
{We investigated the prevalence of giant radio galaxies (GRGs), some of the largest structures powered by supermassive black holes,  within supercluster environments, and the influence of such environments on their properties.
Utilising two large catalogues of superclusters (401) and GRGs (1446), we established the existence of 77 GRGs (5.3\%) residing in 64 superclusters (16\%) within $\rm 0.05 \leq z \leq 0.42$. Among the 77 GRGs found in superclusters, we identified $\sim$\,70\% as residing within galaxy clusters. Within the subset of GRGs not located in superclusters, which constitutes 94.7\% of the sample, a mere 21\% are associated with galaxy clusters, while the remaining majority are situated in sparser environments. We examined the influence of differing environments, such as cluster versus non-cluster and supercluster versus non-supercluster regions, on the size of GRGs, while also exploring the driving factors behind their overall growth. Our findings show that the largest GRGs ($\gtrsim$\,3 Mpc) grow in
underdense environments beyond the confines of dense environments. Moreover, we show that $\sim$ 24\% of 1446 GRGs reside in galaxy clusters. We conclude that GRGs preferentially grow in sparser regions of the cosmic web and have a significantly larger median size. 
Finally, we demonstrate the potential of GRGs as astrophysical probes with specific cases where GRGs, exhibiting polarised emissions and located behind superclusters (acting as natural Faraday screens), were used to estimate magnetic field strengths of the supercluster environment at sub-microgauss levels.}

\keywords{magnetic fields -- galaxies: active -- galaxies: clusters: general -- large-scale structure of Universe -- radio continuum: galaxies}

\maketitle

\defcitealias{Oei2023}{Oei23}
\defcitealias{Sankhyayan2023}{SSC23}

\section{Introduction} \label{sec:intro}
The matter in the Universe is distributed in an intricate network called the cosmic web \citep{Einasto80,Bond96}, composed of clusters,  filaments, sheets, and voids. Nested within these are superclusters (SCls),  vast conglomerates of galaxies and clusters with
higher number densities than the universal average, spanning tens to hundreds of megaparsecs (Mpc) in size \citep[e.g.][]{Lietzen16,Bagchi17}. The densities within these massive structures are not uniform, but are instead influenced by the spatial distribution of their constituent galaxies and clusters \citep{Liivamagi12}.

Recent studies by \citet{Sankhyayan2023} and \citet{EinastoM24} highlight the supercluster environment's influence on the growth of galaxy groups and clusters. These groups and clusters, on Mpc scales, are impacted by the environments of SCls with sizes ranging from tens to hundreds of Mpc. Given the scales involved and the dynamical nature of these superstructures, it stands to reason to investigate whether similarly vast entities on Mpc scales, such as giant radio galaxies (GRGs) with projected linear sizes $\gtrsim$ 0.7~Mpc, are also subject to the influence of their supercluster habitats.

Giant radio galaxies are a subclass of radio sources powered by supermassive black holes (SMBHs). Their large sizes are inferred to result from extensive jets reaching  Mpc scales \citep[for a review, see][]{GRSREV2023}, though these jets are not often detected in radio images, especially in the most common type of GRGs, the FRIIs \citep{Fanaroff1974}. Until recently, the prevalence of GRGs was perceived to be substantially lower compared to typical radio galaxies (normal-sized), and the explanations for their immense size were largely conjectural. The scarcity of a statistically significant sample of GRGs hindered a thorough investigation. Recent advancements in large sky area radio surveys  \citep[e.g. LOFAR][]{LOTSSDR2} have enabled considerable progress in identifying and comprehensively investigating large samples of GRGs \citep[e.g.][]{PDLOTSS,Oei2023}. While the radio properties and broader active galactic nucleus (AGN) characteristics of GRGs have been extensively studied \citep[e.g.][and  references therein]{Konar2004,sagan1,kuzmic19,sagan2}, fewer studies have been conducted on their large-scale environmental properties \citep[e.g.][]{Komberg2009,Lan2021,2024Oei}. One prevailing hypothesis posits that GRGs predominantly grow in environments with a sparser \citep[e.g.][]{mack98} intergalactic medium (IGM). With the recent increment in the availability of large GRG samples, we can now more effectively study the local and broader environments that influence GRGs. Understanding these environments is crucial to unravelling the growth, formation, and evolution of GRGs. The substantial sizes of GRGs are attributed to their powerful jets, which are likely to influence the properties of their host AGNs and galaxies. Consequently, the study of GRGs is crucial for advancing our understanding of AGN feedback mechanisms.

Recently, \citet{Oeiwhim2023} used radio and X-ray data combined with modelling of GRG (NGC~6185) to estimate the temperature of the warm-hot intergalactic medium (WHIM), demonstrating that GRGs can probe the thermodynamics of galaxy groups and beyond. Their results highlight the importance of GRGs as cosmic probes within supercluster environments, marking a significant advancement in their utility.

The potential for SCls to impact the development and characteristics of GRGs forms the basis of our inquiry, bridging the gap between the largest scale structures in the Universe and the largest jets (and their products) powered by SMBHs.
GRGs, owing to their large sizes and luminous radio emission, can serve as tools for exploring other astrophysical processes. The following points highlight the possible pivotal roles of GRG.

Firstly, GRGs, with their large jets and lobes, are crucial for studying galaxy evolution, particularly in supercluster environments; probing the Universe's large-scale structure; and understanding the shape, composition, and evolution of the cosmic web through their interactions with superclusters. Additionally, by mapping the IGM, GRGs can possibly help address the `missing baryon problem'.
Secondly, GRGs are important for mapping magnetic fields. Once identified in SCls, the  extensive lobes of GRGs can be studied through multi-frequency radio observations to measure magnetic field strengths within these structures. The lobes of GRGs, extending through the IGM, offer insights into magnetic field structures across superclusters. Their large-scale radio emissions make them excellent probes for the magnetic environments in these dense regions. Additionally, identifying GRGs behind superclusters enables probing of the magnetic field through the Faraday rotation method.

To address these points, it is crucial first to identify a statistically significant sample of GRGs within (as well as behind) supercluster environments. Thus, under the project SAGAN\footnote{\url{https://sites.google.com/site/anantasakyatta/sagan}} \citep{D17,sagan1}, our  aim is to systematically study the large-scale environment of GRGs, specifically focusing on quantifying the presence of GRGs in the densest and the largest regions of the Universe, such as SCls. 
We  compare the properties of GRGs within superclusters to those outside, examining how their size, radio morphology, and spectral characteristics are influenced by their environment. Factors like their supercluster location can affect their morphology, leading to potential distortions, and may correlate with their current state of activity, whether active or relic (discernible from their radio structure and spectral features). This investigation is vital for revealing how environmental factors impact the lifecycle of GRGs.

The study by \citet{Mauduit2007} is a notable exception in the extremely limited research examining radio galaxies in the context of their supercluster environment and its effects on them. Their study on the Shapley Supercluster \citep{Raychaudhury89} found that the dense core regions, where clusters are merging, have the fewest radio galaxies or sometimes none at all. They suggest that this is likely due to the disruption of their fuel supply (for feeding SMBHs) caused by major mergers between clusters.
When considering GRGs, the findings to date have been  minimal, with only one identified in the Horologium-Reticulum Supercluster \citep{Safouris2009} and two in the Supercluster XLSSsC N03 \citep{Horellou2018}.

Throughout this paper, we  adopted the flat $\Lambda$ cold dark matter ($\Lambda$CDM) cosmology with the following cosmological parameters: $\rm H_0 = 67.66$ km s$^{-1}$ Mpc$^{-1}$, $\rm \Omega_{m0} = 0.3111$, and $\rm \Omega_{\Lambda0} = 0.6889$ \citep{Planck18}.

\section{Giant radio galaxy and supercluster data}
\textbf{Giant radio galaxy data}: We utilised the GRG catalogues of \citet{PDLOTSS} and \citet{Oei2023}, derived from wide-sky area LoFAR Two Metre Sky Survey \citep[LoTSS;][]{LOTSSDR2}, along with GRG catalogues derived from LoFAR deep field surveys \citep{Simonte2022,Simonte2023}. These radio surveys have excellent sensitivity to diffuse radio emission. Given the depth of the surveys and the meticulous validation process, these catalogues can be considered robust and comprehensive.
Since the GRGs reported by all the above-mentioned catalogues are exclusively new discoveries, we   retrieved additional GRGs documented in the literature corresponding to the sky area  covered by LoTSS, through the SAGAN \citep{sagan1} catalogue\footnote{\url{https://vizier.cds.unistra.fr/viz-bin/VizieR-3?-source=J/A\%2bA/642/A153/grs}} as it is a compendium of all GRGs reported till 2020. For our study, we amalgamated all these resources to formulate an enhanced \texttt{GRG-catalogue}. Subsequently, we applied specific redshift and sky area filters to align this
\texttt{GRG-catalogue}
with the supercluster catalogue, thereby facilitating a more targeted and effective analysis of
GRGs within these overdense regions (see Sect.~\ref{sec:analysis}).

\textbf{Supercluster data}: We utilised the supercluster catalogue (\texttt{SCl-catalogue}) of \citet{Sankhyayan2023}  (hereafter SSC23) with 662 SCls. They identified SCls by applying a modified friends of friends (FoF) algorithm on the cluster catalogue of \citet[WH15,][]{Wen15},
which is based on the SDSS DR12 \citep{Alam15}. They found that in a supercluster environment, member clusters are more massive (high richness) compared to clusters in a non-supercluster environment, indicating the influence of the supercluster environment on their evolution.
They showed that a supercluster has a gravitational influence on its constituents visible in the phase-space distribution of the mock SCls extracted from the Horizon Run 4 \citep[HR4,][]{Kim15} simulation. In this catalogue, the median mass of SCls is $\sim 5.8 \times 10^{15}$ M$_{\odot}$ and the median size is $\sim 65$ Mpc (\citetalias{Sankhyayan2023}).
The advantage of using the catalogue of \citetalias{Sankhyayan2023} is its wide sky-area coverage ($\sim$ 14000 deg$^2$) and relatively large redshift span ($0.05 \leq z \leq 0.42$). In addition, this catalogue also provides the masses and density contrast ($\delta$) of SCls.
In contrast to other catalogues \citep[e.g.][]{Liu2024} that classify even cluster pairs as superclusters, \citetalias{Sankhyayan2023} identifies superclusters with at least ten member clusters. This is made feasible by the comprehensiveness of the WH15 catalogue. Defining supercluster boundaries in 3D space with only two to three member clusters is very challenging, as is identifying GRGs within them.

\section{Identifying giant radio galaxies in superclusters and clusters}
\label{sec:analysis}
Ensuring a fair cross-matching between the \texttt{GRG-catalogue} and the \texttt{SCl-catalogue} requires finding the common sky coverage and redshift span in the two datasets. Moreover, for the purpose of investigating potential distinctions in the properties of SCls hosting GRGs compared to those without any GRGs (see Sect.~\ref{sec:r1}), it is crucial to exclude sky regions containing SCls but without the sky coverage of the \texttt{GRG-catalogue}. This process involves exclusively considering GRGs within the sky-coverage mask defined by the input cluster catalogue used to make the \texttt{SCl-catalogue}, as detailed in \citetalias{Sankhyayan2023}. The sky mask is generated using a \texttt{Healpix} map with a resolution parameter of $N_{side} = 64$. Pixels harbouring at least one cluster are assigned a value of 1, while those devoid of  clusters receive a value of 0. A similar procedure is employed to generate a sky mask for the \texttt{GRG-catalogue}; however, a reduced resolution of $N_{side} = 16$ is used to accommodate the lower number density of GRGs on the sky-plane compared to clusters. The intersection of these two masks defines the region of interest for our study (see Fig.~\ref{fig:sky_mask}). Additionally, a redshift constraint of $0.05 \leq z \leq 0.42$ is implemented on the \texttt{GRG-catalogue} to align its redshift span with that of the \texttt{SCl-catalogue}. These constraints give a \texttt{GRG-sample} of 1446 and a \texttt{SCl-sample} of 401 SCls.

Given the complex geometries of SCls, delineating their boundaries presents a significant challenge. This challenge is heightened when using the FoF algorithm for supercluster identification as it provides discrete locations for member clusters. Using an ellipsoid fit to define the supercluster boundary may overestimate its volume, while a wrapping method risks underestimation due to the dependence  on a parameter determining the extent of wrapping. To strike a balance, we adopted the convex hull of member clusters as the supercluster boundary.

\begin{figure}
  \centering
  \includegraphics[width=\linewidth]{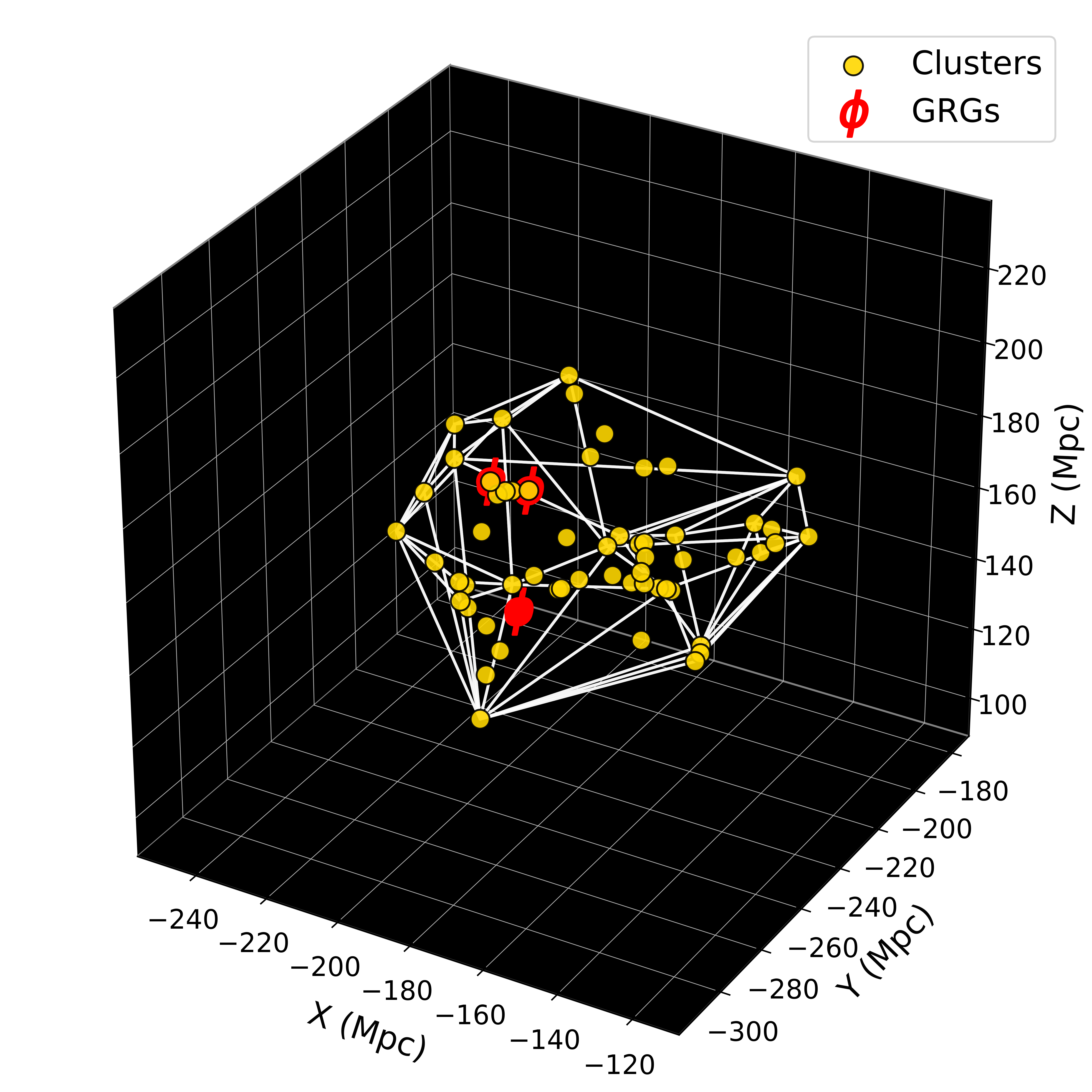}
  \caption{Three GRGs (red markers) inside SCl~2.
  Two of the GRGs are the brightest cluster galaxies.
  Here, the supercluster SCl~2 is represented by convex hull edges (white lines), encompassing its 57 member galaxy clusters shown in yellow (see Sect.~\ref{sec:analysis} and Table~\ref{tab:oeigrg} for more details).  X, Y, and Z are the equatorial comoving coordinates.}
  \label{fig:conhul}
\end{figure}

In 3D Euclidean space, the convex hull of a set of points \textbf{P} represents the minimum-volume convex surface encompassing \textbf{P} \citep{Barber96}.
To locate GRGs within a supercluster boundary, we cross-matched the \texttt{GRG-sample} with the \texttt{SCl-sample}, identifying GRGs within the convex hull region formed by the member clusters of a supercluster (see Fig.~\ref{fig:conhul} for an example).
\citetalias{Sankhyayan2023} also used the convex hull of member clusters to estimate the SCl volumes and subsequently calculated their densities (or density contrast). This approach supports the use of a convex hull for boundary estimation in this study, particularly when investigating the properties of SCls hosting GRGs compared to those without GRGs.

We also located GRGs within or outside galaxy clusters (Cls) using the cluster catalogue of \citet{Wen15}, while also suppressing the fingers-of-god redshift distortion induced by peculiar velocities of galaxies within clusters \citep{Jackson1972}. First, we identified all GRGs in the \texttt{GRG-sample}, irrespective of their SCl environment, within the projected cluster radius on the sky plane and falling within a certain redshift range determined by the cluster's velocity dispersion. Then, these GRGs were labelled as being in the Cl environment, and their fingers-of-god redshift distortion was suppressed accordingly. The details of this method are presented in Appendix~\ref{sec:clust_mem}.

\section{Results: The influence of environment on the sizes of giant radio galaxies} \label{sec:r1}
We used the \texttt{GRG-sample} of 1446 GRGs for the analysis of their large-scale environments. As mentioned in Sect.~\ref{sec:analysis}, we  also identified GRGs that are members of clusters  both inside and outside of SCl environments. We found that $\sim$24\% GRGs are in galaxy clusters (see Table~\ref{tab:grg_env}), which is similar to the finding of  \citet{2024Oei} for lower redshifts ($z<$~0.16). Our analysis shows that GRGs associated with galaxy clusters have a median size of 1.1 Mpc, regardless of their supercluster environment, while those in non-cluster environments have a slightly larger median size of 1.2 Mpc. We conducted a non-parametric Mann-Whitney U test to compare these sample medians (W = 171757, p = 0.007), indicating a significant difference in the median sizes of GRGs between cluster and non-cluster environments.

\begin{figure}
\centering
\includegraphics[width=\columnwidth]{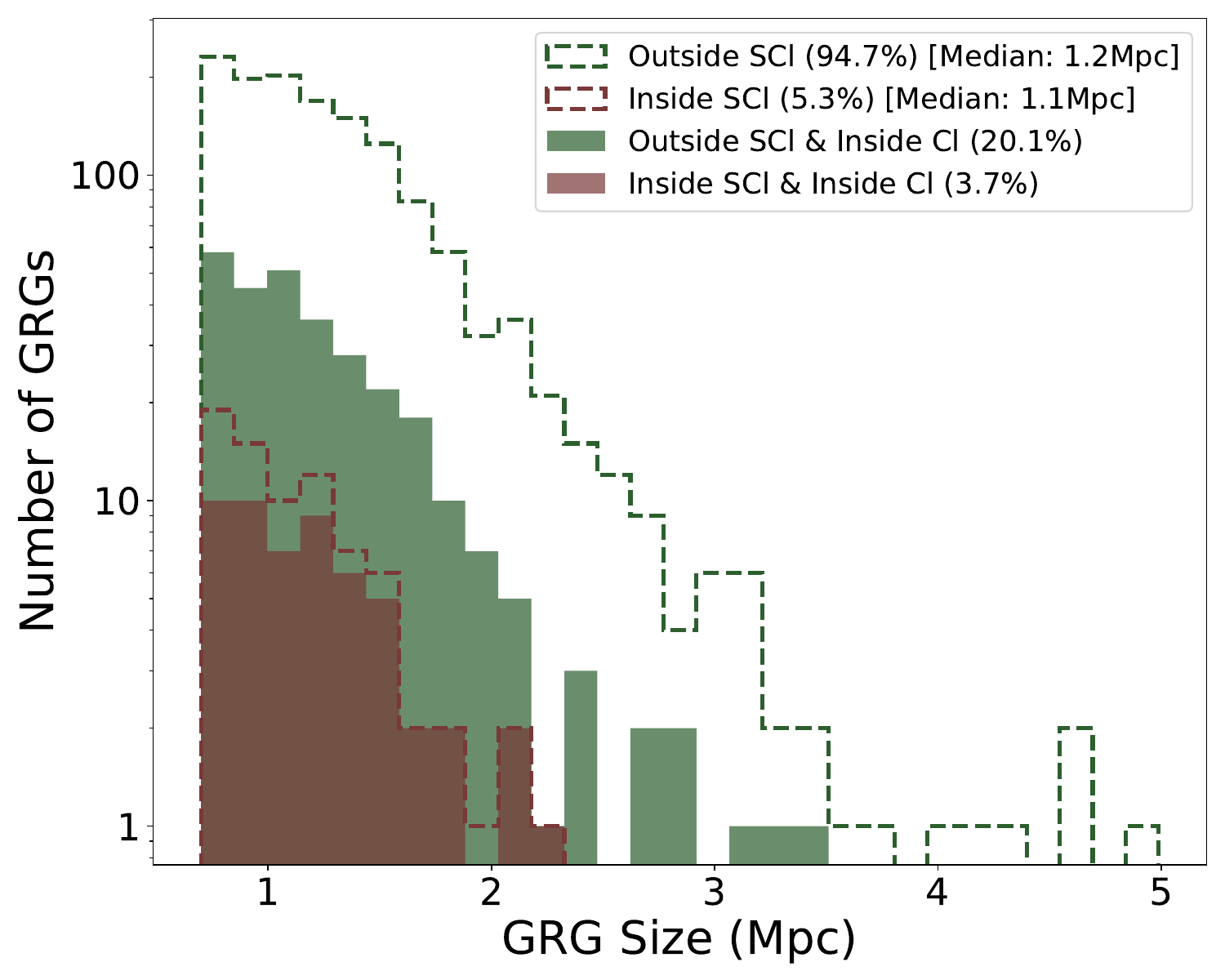}
\caption{Size distribution of GRGs. The green dashed line shows GRGs outside superclusters;  the green solid histogram represents those outside superclusters but within clusters. The brown dashed line indicates GRGs within superclusters, and the brown solid histogram shows GRGs within both superclusters and clusters. The percentages and median sizes of GRGs inside and outside SCls are also shown in the legend.}
\label{fig:grgsize}
\end{figure}

The distribution of GRGs and their sizes from the \texttt{GRG-sample} across various environments is presented in Table~\ref{tab:grg_env} and Fig.~\ref{fig:grgsize}, respectively, to assess which favours their growth. Below, we present our key findings:

\begin{itemize}

    \item Approximately 95\% of GRGs\footnote{The relevant data tables of GRGs in non-supercluster environments, including their cluster and non-cluster associations, will be made available through VizieR.} are located outside supercluster environments. Fig.~\ref{fig:grgsize} clearly shows that the largest ($\gtrsim$\,3~Mpc) GRGs grow in relatively isolated environments beyond the confines of SCls or galaxy clusters. The majority of the GRGs from the \texttt{GRG-sample} are outside SCls, and of these $\sim$\,21\% evidently grow in galaxy clusters (see Table~\ref{tab:grg_env}).

\item Using the information about the extent and overall geometry of superclusters and their constituent member galaxy clusters, we present for the first time the distribution of GRGs in the complex architecture of SCls (see Fig.~\ref{fig:conhul}).
A mere 5.3\% (77 GRGs) of GRGs, with a median size of $\sim$\,1.1 Mpc, are located within superclusters (listed in Table~\ref{tab:oeigrg}), suggesting that supercluster environments likely impose a restrictive influence on the growth of GRGs, as indicated by   their limited presence and   smaller median size (see Fig.~\ref{fig:grgsize}). In this subset of GRGs found in SCls,
$\sim$70\% (47 BCGs and 7 galaxy cluster members) reside in galaxy clusters (see Col. 13 of Table~\ref{tab:oeigrg}) and 23 are located outside of galaxy clusters. 

\end{itemize}

\subsection{Variations in giant radio galaxy properties: A statistical analysis of supercluster associations}

To probe for variations in GRG attributes contingent on their supercluster association, we conducted a two-sample Kolmogorov-Smirnov (KS) test.
This analysis focused on three key aspects: the size distribution of GRGs inside versus outside SCls, the mass of the central SMBHs within these GRGs, and the density contrasts of SCls that host GRGs compared to those that do not. The KS test evaluates the null hypothesis that the two samples originate from identical distributions. The null hypothesis is rejected if the p-value of the KS test is $> 0.05$. The parameters of the KS test are presented in Table~\ref{tab:ks}. Here we discuss the outcomes of the test and their implications.

We observe significant size differences between GRGs within and outside SCls, which is also seen in Fig.~\ref{fig:grgsize} (distributions represented by dashed green and dashed brown lines). We observe that GRGs within the confines of SCls, particularly those in Cl environments, predominantly span sizes up to $\sim$\,1.5~Mpc, with an absence of GRGs exceeding $\sim$\,2.5 Mpc. Conversely, GRGs outside of SCl boundaries exhibit a potential for greater expansion, reaching sizes up to $\sim$\,5~Mpc. The distributions conclusively show the constraining effect of SCl environments on the growth of GRGs to larger sizes ($\gtrsim$ 2~Mpc).

We compared the masses of SMBHs hosting GRGs located both within SCls (57/77 GRGs, as per availability of SMBH masses in \citealt{Oei2023}) and outside  SCls (683/1369 GRGs).
The KS test's p-value of 0.08 (see Table~\ref{tab:ks}) at best suggests marginal evidence that these are drawn from different populations.
The median values of the masses of SMBHs inside and outside SCls are $\rm 1.05 \times 10^9~M_{\odot}$ and $\rm 9.24 \times 10^8~M_{\odot}$, respectively.

\begin{figure}
\centering
\includegraphics[width=\columnwidth]{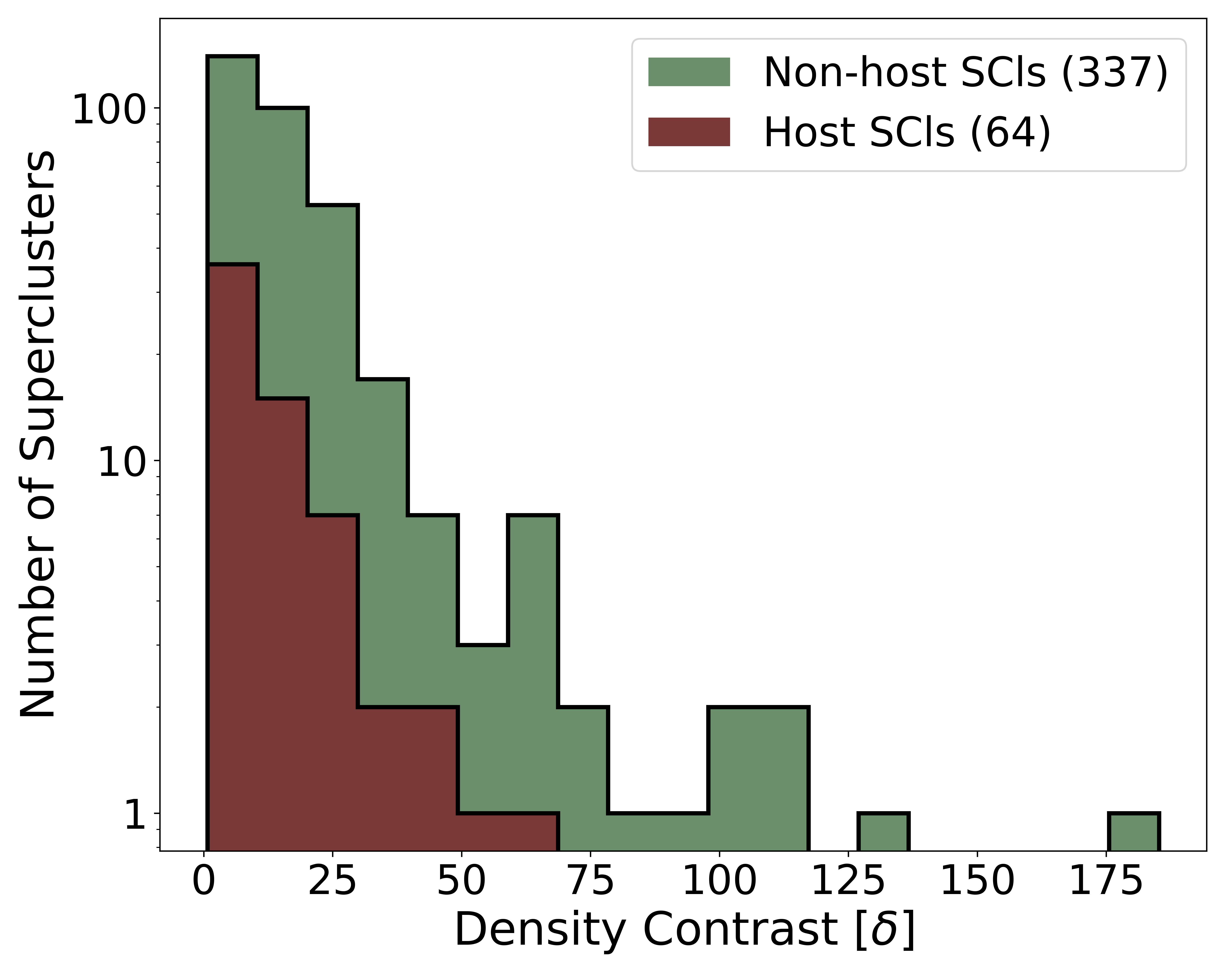}
\caption{
Distribution of density contrast of SCls hosting GRGs (64)  compared to those that do not host GRGs (337). It indicates that superclusters with high densities impede the growth of a radio galaxy, preventing it from developing into a GRG.}
\label{fig:super_density}
\end{figure}

Fig.~\ref{fig:super_density} shows the distribution of density contrast ($\rm \delta$) between superclusters that contain GRGs (red histogram) and those that do not host any GRGs (green histogram). It shows that the location of GRGs is preferentially skewed towards lower-density SCls (also see Table~\ref{tab:ks}). The presence of GRGs in SCls with a density contrast $\delta \gtrsim 25$ is markedly rare (only six).
Among the 401 superclusters analyzed, only 64 (16\%) contain GRGs. This shows that the majority of these large coherent overdense regions within the cosmic web do not facilitate the growth of GRGs.

Our analysis (see Figs.~\ref{fig:grgsize} and \ref{fig:localisation}) also suggests that the densest regions within SCls, particularly those close to massive and nearby member clusters, are not the most favourable for the formation of the largest GRGs ($\gtrsim$3~Mpc). These regions could be characterised by more complex gravitational interactions, potentially leading to disruptions in the radio jet propagation that is necessary for GRG growth. Our visual inspection of radio images from LoTSS DR2 data reveals that fewer than 10\% of GRGs within SCls do not show major structural distortions, such as bent lobes or high asymmetry.

\begin{table}
  \centering
  \caption{Percentage of 1446 GRGs from the \texttt{GRG-sample} in different environments.}
  \label{tab:grg_env}
  \begin{tabular}{l c c}
    \hline
    \textit{\textbf{Environment}} & \textbf{Cluster} & \textbf{Non-Cluster}\\
    \hline
    \textbf{Supercluster} & 3.7\% & 1.6\% \\
    \textbf{Non-Supercluster} & 20.1\% & 74.6\% \\
    \hline
  \end{tabular}
\end{table}

\begin{table}
  \centering
  \caption{Detailed statistics of the KS test. More details in Sec.~\ref{sec:r1}.}
  \label{tab:ks}
  \resizebox{\columnwidth}{!}{
  \begin{tabular}{l c c}
    \hline
    \textbf{Samples} & \textbf{Statistic} & \textbf{p-value} \\
    \hline
    GRGs Size (inside and outside SCl) & 0.17 & 0.03 \\
    GRGs SMBH Mass (inside and outside SCl) & 0.17 & 0.08 \\
    SCls $\delta$ (host and non-host SCl) & 0.20 & 0.02 \\
    \hline
  \end{tabular}
  }
\end{table}

\subsection{Two exceptional superclusters with giant radio galaxies}
We identified two exceptional superclusters, SCl~2 and SCl~73, hosting three GRGs (see Fig.~\ref{fig:conhul} and Table~\ref{tab:oeigrg}). Notably, SCl~2 is the second most massive ($\rm M \approx  2.55\times 10^{16}~\mathrm{M}_{\sun}$) supercluster at $z \approx 0.08$ listed in \citetalias{Sankhyayan2023}, making the presence of three GRGs in this dense environment remarkable.
It extends to $\approx 124$ Mpc and contains 57 member clusters. SCl 2 is a composite entity of two substantial sub-structures, the  Corona Borealis (CB) supercluster and the  Abell 2142 supercluster,  weakly connected by a chain of galaxies and poor groups \citep{Pillastrini2019, EinastoM21}. Two of the three GRGs in  SCl~2, which are also BCGs, exhibit bent radio morphologies, underscoring the influence of their dynamic local environment. It is often found that the dominant galaxies in clusters or groups are  wide-angle tail (WAT) radio sources, known for their powerful, bent emissions \citep{2023WATREV}.
Similarly, we find three GRGs in SCl~73, which is located at $z \approx 0.14$, has a  mass $\rm M \approx 1.02 \times 10^{16} \mathrm{M}_{\sun}$, has a spatial extent of $\approx 94$ Mpc, and encompasses a total of 21 member galaxy clusters.
Furthermore, there are nine other SCls with two GRGs in each (see Table~\ref{tab:oeigrg} for details).

\section{Giant radio galaxies as probes of the magnetic field in superclusters} \label{sec:grgprobe}
As described in Sect.~\ref{sec:intro}, the objective of identifying GRGs within supercluster environments extends beyond examining the influence of these dense and massive structures on the GRGs themselves. GRGs serve a dual purpose in this context: they are the subjects of study in terms of how extreme environments impact their morphology and evolution, and they are also instrumental as astrophysical probes. It can be inferred from studies \citep[e.g.][and references therein]{Strom1973,OSullivan2018,Sullivan2023} that large radio galaxies like GRGs show a higher degree of polarisation at low radio frequencies, and hence they can more effectively be used as probes with their Faraday rotation measurements.

We propose here that  GRGs offer a unique vantage point when positioned behind SCls. Utilising linear radio polarisation and Faraday rotation measurements, they become effective tools for constraining the intra-supercluster medium's magnetic field and thermal electron density (see    Fig.~\ref{fig:supeillus}). This methodology has been successfully applied to filaments in the cosmic web \citep[e.g.][]{Sullivan2019}. Building on this, the recent Faraday rotation measurements (RMs) from \citet{Sullivan2023} allow us to infer the magnetic field strength of the supercluster. Since these measurements are based on low-frequency observations, they offer higher precision in RM values and are also more sensitive to weaker magnetic fields, such as those pervasive, possibly in the supercluster environment.
From our \texttt{GRG-sample}, we reliably identified eight GRGs located behind SCls with available RM measurements from \citet{Sullivan2023}, as listed in Table~\ref{tab:magfield}. We only included GRGs with spectroscopic redshift measurements, non-cluster members, and non-supercluster members (see Sect.~\ref{sec:analysis}) to minimize ambiguity and to mitigate the local contribution to the RM values.
The polarised emission from the GRGs probes or samples a substantial region ($\Delta$L$_{\perp,\mathrm{SCl}}$) of the supercluster on the sky plane. The path length ($\Delta$L$_{\parallel,\mathrm{SCl}}$) within the supercluster, covered by the polarised radio waves from the background GRGs, is estimated by taking the redshift limits ($\Delta$z) of the supercluster (i.e. the extent of the supercluster along the direction of the line of sight).
Using the NASA Extragalactic Database (NED), we also verified that the polarised emission traverses the intercluster magneto-ionic medium of the supercluster (responsible for the Faraday rotation) and not through any known core regions of galaxy clusters.
By assuming negligible background\footnote{In our analysis, we assume that contributions to RM from effects or processes local to the GRG are negligible (unless it is in a Cl or SCl). The possible minor local contribution to RM may occur due to various local environment factors as discussed in \citet{Rudnick2003}.} and foreground effects on RM and adopting electron density (n$_{\rm e}$) values of 10$^{-6}$ and 10$^{-7}$ cm$^{-3}$ (reasonable for the intra-SCl environment), we obtained the magnetic field strengths of  sub-$\muup$G (for further details, see Appendix~\ref{sec:appendixA}) and these are averaged on scales of $\Delta$L$_{\perp,\mathrm{SCl}}$ given in Col. 12 of Table~\ref{tab:magfield}. These values are comparable to those obtained by \citet{xu2006} using similar techniques.
Furthermore, using the equipartition method, \citet{Venturi2022} estimated a magnetic field strength of 0.78~$\muup$G over a scale of $\sim$\,1 Mpc in the Shapley supercluster, which is in a range similar to ours.

In summary, we present a compelling proof of concept for employing GRGs as probes to determine the magnetic field strengths within SCls, utilising the Faraday rotation measure technique. This novel approach sets the stage for future work, wherein there exists the potential to comprehensively map the magnetic field distributions across entire SCls, bolstered by improved statistics \citep[e.g. using POSSUM][]{2024possum}.

\section{Summary}
We have presented a comprehensive examination of large-scale environments of GRGs in superclusters and galaxy clusters. We have shown that for the redshift range of $0.05 \leq z \leq 0.42$, only $\sim$16\% of superclusters (of only moderate densities) host GRGs, which is only about 5.3\% of the \texttt{GRG-sample}, clearly showing that the sparser cosmic web environments favour the growth of GRGs. This is further substantiated by the fact that the largest GRGs ($\gtrsim$ 3~Mpc) are found outside of clusters or superclusters. Our analysis indicates that GRGs in cluster environments are   smaller than those in non-cluster environments, regardless of whether the clusters are part of a supercluster.
Furthermore, GRGs in cluster environments can grow to sizes $\gtrsim$ 2 Mpc if the host cluster is not part of a supercluster. However, within superclusters, cluster-GRGs rarely exceed 2 Mpc, highlighting the restrictive influence of supercluster environments.

In non-supercluster environments most GRGs are non-BCGs, whereas in superclusters the majority are BCGs. This finding contrasts with the typical preference of GRGs for sparser environments, highlighting a unique characteristic within superclusters. However, sparse environments are not the sole factor for GRG growth to megaparsec scales; jet kinetic power, dependent on the AGN's power and accretion state, also plays a crucial role. BCGs, particularly in superclusters, having grown through multiple mergers, are likely to host the most powerful AGNs capable of producing jets that pierce through the dense intracluster medium and intra-supercluster medium, thereby forming GRGs. Additionally, BCGs are more likely to exhibit extended radio emission compared to other cluster members \citep[e.g.][]{Garon2019}, increasing their probability of being GRGs.
Also, our findings conclusively show that a majority ($\sim$76\% of \texttt{GRG-sample}) of GRGs reside in non-cluster environments, like filaments and voids, which will be further investigated in our future work as part of the SAGAN project.

Moreover, this study has uniquely identified eight GRGs (with RM information) directly behind six superclusters on the sky plane for the first time. Using RM data, we demonstrated their potential to estimate the magnetic field strength in the supercluster's intercluster medium. This investigation lays the groundwork for further characterising GRGs in various environments and establishes GRGs as powerful probes for exploring astrophysical phenomena, including the magnetised cosmic web.

Overall, this work represents an initial step in identifying the large-scale environments of GRGs and their effects using the best available catalogues of GRGs, galaxy clusters, and superclusters in a statistical manner. Future studies can build on this foundation with improved data, and more detailed investigations will be possible with upcoming survey data from DESI \citep{desiEDR2023}, \textit{Euclid} \citep{Euclid}, and 4MOST \citep{de_Jong19}, enhancing the robustness of our findings.

\begin{acknowledgements}
We thank the referee for detailed and useful comments on the manuscript.
We dedicate this paper to the loving memory of SS's late grandfather (Dr. Narendra Sankhyayan).
We also thank Elmo Tempel, Shabbir Sheikh, Maret Einasto and Vikram Khaire for their valuable comments on the manuscript.
SS acknowledges the support of the ETAg grant PRG1006 and the HTM project ``Foundations of the Universe'' (TK202).
PD acknowledges support from the Spanish Ministry of Science \& Innovation under the grant -``PARSEC: Multiwavelength investigations of the central parsec of galaxies" (PID2020-114092GB-I00). The NASA/IPAC Extragalactic Database (NED) is funded by the National Aeronautics and Space Administration and operated by the California Institute of Technology.
\end{acknowledgements}

\bibliographystyle{aa} 
\bibliography{SAGAN_IV}

\begin{thebibliography}{52}
\expandafter\ifx\csname natexlab\endcsname\relax\def\natexlab#1{#1}\fi

\bibitem[{{Alam} {et~al.}(2015){Alam}, {Albareti}, {Allende Prieto}, {Anders},
  {Anderson}, {Anderton}, {Andrews}, {Armengaud}, {Aubourg}, {Bailey}, \&
  et~al.}]{Alam15}
{Alam}, S., {Albareti}, F.~D., {Allende Prieto}, C., {et~al.} 2015, \apjs, 219,
  12

\bibitem[{{Bagchi} {et~al.}(2017){Bagchi}, {Sankhyayan}, {Sarkar},
  {Raychaudhury}, {Jacob}, \& {Dabhade}}]{Bagchi17}
{Bagchi}, J., {Sankhyayan}, S., {Sarkar}, P., {et~al.} 2017, \apj, 844, 25

\bibitem[{Barber {et~al.}(1996)Barber, Dobkin, \& Huhdanpaa}]{Barber96}
Barber, C.~B., Dobkin, D.~P., \& Huhdanpaa, H. 1996, ACM Trans. Math. Softw.,
  22, 469

\bibitem[{{Bond} {et~al.}(1996){Bond}, {Kofman}, \& {Pogosyan}}]{Bond96}
{Bond}, J.~R., {Kofman}, L., \& {Pogosyan}, D. 1996, \nat, 380, 603

\bibitem[{{Dabhade} {et~al.}(2020{\natexlab{a}}){Dabhade}, {Combes},
  {Salom{\'e}}, {Bagchi}, \& {Mahato}}]{sagan2}
{Dabhade}, P., {Combes}, F., {Salom{\'e}}, P., {Bagchi}, J., \& {Mahato}, M.
  2020{\natexlab{a}}, \aap, 643, A111

\bibitem[{{Dabhade} {et~al.}(2017){Dabhade}, {Gaikwad}, {Bagchi},
  {Pandey-Pommier}, {Sankhyayan}, \& {Raychaudhury}}]{D17}
{Dabhade}, P., {Gaikwad}, M., {Bagchi}, J., {et~al.} 2017, \mnras, 469, 2886

\bibitem[{{Dabhade} {et~al.}(2020{\natexlab{b}}){Dabhade}, {Mahato}, {Bagchi},
  {Saikia}, {Combes}, {Sankhyayan}, {R{\"o}ttgering}, {Ho}, {Gaikwad},
  {Raychaudhury}, {Vaidya}, \& {Guiderdoni}}]{sagan1}
{Dabhade}, P., {Mahato}, M., {Bagchi}, J., {et~al.} 2020{\natexlab{b}}, \aap,
  642, A153

\bibitem[{{Dabhade} {et~al.}(2020{\natexlab{c}}){Dabhade}, {R{\"o}ttgering},
  {Bagchi}, {Shimwell}, {Hardcastle}, {Sankhyayan}, {Morganti}, {Jamrozy},
  {Shulevski}, \& {Duncan}}]{PDLOTSS}
{Dabhade}, P., {R{\"o}ttgering}, H.~J.~A., {Bagchi}, J., {et~al.}
  2020{\natexlab{c}}, \aap, 635, A5

\bibitem[{{Dabhade} {et~al.}(2023){Dabhade}, {Saikia}, \&
  {Mahato}}]{GRSREV2023}
{Dabhade}, P., {Saikia}, D.~J., \& {Mahato}, M. 2023, Journal of Astrophysics
  and Astronomy, 44, 13

\bibitem[{{de Jong} {et~al.}(2019){de Jong}, {Agertz}, {Berbel}, {Aird},
  {Alexander}, {Amarsi}, {Anders}, {Andrae}, {Ansarinejad}, {Ansorge},
  {Antilogus}, {Anwand-Heerwart}, {Arentsen}, {Arnadottir}, {Asplund}, {Auger},
  {Azais}, {Baade}, {Baker}, {Baker}, {Balbinot}, {Baldry}, {Banerji},
  {Barden}, {Barklem}, {Barth{\'e}l{\'e}my-Mazot}, {Battistini}, {Bauer},
  {Bell}, {Bellido-Tirado}, {Bellstedt}, {Belokurov}, {Bensby}, {Bergemann},
  {Bestenlehner}, {Bielby}, {Bilicki}, {Blake}, {Bland-Hawthorn}, {Boeche},
  {Boland}, {Boller}, {Bongard}, {Bongiorno}, {Bonifacio}, {Boudon}, {Brooks},
  {Brown}, {Brown}, {Br{\"u}ggen}, {Brynnel}, {Brzeski}, {Buchert},
  {Buschkamp}, {Caffau}, {Caillier}, {Carrick}, {Casagrande}, {Case}, {Casey},
  {Cesarini}, {Cescutti}, {Chapuis}, {Chiappini}, {Childress}, {Christlieb},
  {Church}, {Cioni}, {Cluver}, {Colless}, {Collett}, {Comparat}, {Cooper},
  {Couch}, {Courbin}, {Croom}, {Croton}, {Daguis{\'e}}, {Dalton}, {Davies},
  {Davis}, {de Laverny}, {Deason}, {Dionies}, {Disseau}, {Doel}, {D{\"o}scher},
  {Driver}, {Dwelly}, {Eckert}, {Edge}, {Edvardsson}, {Youssoufi}, {Elhaddad},
  {Enke}, {Erfanianfar}, {Farrell}, {Fechner}, {Feiz}, {Feltzing}, {Ferreras},
  {Feuerstein}, {Feuillet}, {Finoguenov}, {Ford}, {Fotopoulou}, {Fouesneau},
  {Frenk}, {Frey}, {Gaessler}, {Geier}, {Gentile Fusillo}, {Gerhard},
  {Giannantonio}, {Giannone}, {Gibson}, {Gillingham},
  {Gonz{\'a}lez-Fern{\'a}ndez}, {Gonzalez-Solares}, {Gottloeber}, {Gould},
  {Grebel}, {Gueguen}, {Guiglion}, {Haehnelt}, {Hahn}, {Hansen}, {Hartman},
  {Hauptner}, {Hawkins}, {Haynes}, {Haynes}, {Heiter}, {Helmi}, {Aguayo},
  {Hewett}, {Hinton}, {Hobbs}, {Hoenig}, {Hofman}, {Hook}, {Hopgood},
  {Hopkins}, {Hourihane}, {Howes}, {Howlett}, {Huet}, {Irwin}, {Iwert},
  {Jablonka}, {Jahn}, {Jahnke}, {Jarno}, {Jin}, {Jofre}, {Johl}, {Jones},
  {J{\"o}nsson}, {Jordan}, {Karovicova}, {Khalatyan}, {Kelz}, {Kennicutt},
  {King}, {Kitaura}, {Klar}, {Klauser}, {Kneib}, {Koch}, {Koposov},
  {Kordopatis}, {Korn}, {Kosmalski}, {Kotak}, {Kovalev}, {Kreckel}, {Kripak},
  {Krumpe}, {Kuijken}, {Kunder}, {Kushniruk}, {Lam}, {Lamer}, {Laurent},
  {Lawrence}, {Lehmitz}, {Lemasle}, {Lewis}, {Li}, {Lidman}, {Lind}, {Liske},
  {Lizon}, {Loveday}, {Ludwig}, {McDermid}, {Maguire}, {Mainieri}, {Mali},
  {Mandel}, {Mandel}, {Mannering}, {Martell}, {Martinez Delgado}, {Matijevic},
  {McGregor}, {McMahon}, {McMillan}, {Mena}, {Merloni}, {Meyer}, {Michel},
  {Micheva}, {Migniau}, {Minchev}, {Monari}, {Muller}, {Murphy},
  {Muthukrishna}, {Nandra}, {Navarro}, {Ness}, {Nichani}, {Nichol}, {Nicklas},
  {Niederhofer}, {Norberg}, {Obreschkow}, {Oliver}, {Owers}, {Pai},
  {Pankratow}, {Parkinson}, {Paschke}, {Paterson}, {Pecontal}, {Parry},
  {Phillips}, {Pillepich}, {Pinard}, {Pirard}, {Piskunov}, {Plank},
  {Pl{\"u}schke}, {Pons}, {Popesso}, {Power}, {Pragt}, {Pramskiy}, {Pryer},
  {Quattri}, {Queiroz}, {Quirrenbach}, {Rahurkar}, {Raichoor}, {Ramstedt},
  {Rau}, {Recio-Blanco}, {Reiss}, {Renaud}, {Revaz}, {Rhode}, {Richard},
  {Richter}, {Rix}, {Robotham}, {Roelfsema}, {Romaniello}, {Rosario},
  {Rothmaier}, {Roukema}, {Ruchti}, {Rupprecht}, {Rybizki}, {Ryde}, {Saar},
  {Sadler}, {Sahl{\'e}n}, {Salvato}, {Sassolas}, {Saunders}, {Saviauk},
  {Sbordone}, {Schmidt}, {Schnurr}, {Scholz}, {Schwope}, {Seifert}, {Shanks},
  {Sheinis}, {Sivov}, {Sk{\'u}lad{\'o}ttir}, {Smartt}, {Smedley}, {Smith},
  {Smith}, {Sorce}, {Spitler}, {Starkenburg}, {Steinmetz}, {Stilz}, {Storm},
  {Sullivan}, {Sutherland}, {Swann}, {Tamone}, {Taylor}, {Teillon}, {Tempel},
  {ter Horst}, {Thi}, {Tolstoy}, {Trager}, {Traven}, {Tremblay}, {Tresse},
  {Valentini}, {van de Weygaert}, {van den Ancker}, {Veljanoski}, {Venkatesan},
  {Wagner}, {Wagner}, {Walcher}, {Waller}, {Walton}, {Wang}, {Winkler},
  {Wisotzki}, {Worley}, {Worseck}, {Xiang}, {Xu}, {Yong}, {Zhao}, {Zheng},
  {Zscheyge}, \& {Zucker}}]{de_Jong19}
{de Jong}, R.~S., {Agertz}, O., {Berbel}, A.~A., {et~al.} 2019, The Messenger,
  175, 3

\bibitem[{{DESI Collaboration} {et~al.}(2023){DESI Collaboration}, {Adame},
  {Aguilar}, {Ahlen}, {Alam}, {Aldering}, {Alexander}, {Alfarsy}, {Allende
  Prieto}, {Alvarez}, {Alves}, {Anand}, {Andrade-Oliveira}, {Armengaud},
  {Asorey}, {Avila}, {Aviles}, {Bailey}, {Balaguera-Antol{\'\i}nez},
  {Ballester}, {Baltay}, {Bault}, {Bautista}, {Behera}, {Beltran}, {BenZvi},
  {Beraldo e Silva}, {Bermejo-Climent}, {Berti}, {Besuner}, {Beutler},
  {Bianchi}, {Blake}, {Blum}, {Bolton}, {Brieden}, {Brodzeller}, {Brooks},
  {Brown}, {Buckley-Geer}, {Burtin}, {Cabayol-Garcia}, {Cai}, {Canning},
  {Cardiel-Sas}, {Carnero Rosell}, {Castander}, {Cervantes-Cota}, {Chabanier},
  {Chaussidon}, {Chaves-Montero}, {Chen}, {Chuang}, {Claybaugh}, {Cole},
  {Cooper}, {Cuceu}, {Davis}, {Dawson}, {de Belsunce}, {de la Cruz}, {de la
  Macorra}, {de Mattia}, {Demina}, {Demirbozan}, {DeRose}, {Dey}, {Dey},
  {Dhungana}, {Ding}, {Ding}, {Doel}, {Doshi}, {Douglass}, {Edge},
  {Eftekharzadeh}, {Eisenstein}, {Elliott}, {Escoffier}, {Fagrelius}, {Fan},
  {Fanning}, {Fawcett}, {Ferraro}, {Ereza}, {Flaugher}, {Font-Ribera},
  {Forero-S{\'a}nchez}, {Forero-Romero}, {Frenk}, {G{\"a}nsicke},
  {Garc{\'\i}a}, {Garc{\'\i}a-Bellido}, {Garcia-Quintero}, {Garrison},
  {Gil-Mar{\'\i}n}, {Golden-Marx}, {Gontcho}, {Gonzalez-Morales},
  {Gonzalez-Perez}, {Gordon}, {Graur}, {Green}, {Gruen}, {Guy}, {Hadzhiyska},
  {Hahn}, {Han}, {Hanif}, {Herrera-Alcantar}, {Honscheid}, {Hou}, {Howlett},
  {Huterer}, {Ir{\v{s}}i{\v{c}}}, {Ishak}, {Jacques}, {Jana}, {Jiang},
  {Jimenez}, {Jing}, {Joudaki}, {Jullo}, {Juneau}, {Kizhuprakkat},
  {Kara{\c{c}}ayl{\i}}, {Karim}, {Kehoe}, {Kent}, {Khederlarian}, {Kim},
  {Kirkby}, {Kisner}, {Kitaura}, {Kneib}, {Koposov}, {Kov{\'a}cs}, {Kremin},
  {Krolewski}, {L'Huillier}, {Lambert}, {Lamman}, {Lan}, {Landriau}, {Lang},
  {Lange}, {Lasker}, {Le Guillou}, {Leauthaud}, {Levi}, {Li}, {Linder},
  {Lyons}, {Magneville}, {Manera}, {Manser}, {Margala}, {Martini}, {McDonald},
  {Medina}, {Medina-Varela}, {Meisner}, {Mena-Fern{\'a}ndez}, {Meneses-Rizo},
  {Mezcua}, {Miquel}, {Montero-Camacho}, {Moon}, {Moore}, {Moustakas},
  {Mueller}, {Mundet}, {Mu{\~n}oz-Guti{\'e}rrez}, {Myers}, {Nadathur},
  {Napolitano}, {Neveux}, {Newman}, {Nie}, {Nikutta}, {Niz}, {Norberg},
  {Noriega}, {Paillas}, {Palanque-Delabrouille}, {Palmese}, {Zhiwei},
  {Parkinson}, {Penmetsa}, {Percival}, {P{\'e}rez-Fern{\'a}ndez},
  {P{\'e}rez-R{\`a}fols}, {Pieri}, {Poppett}, {Porredon}, {Pothier}, {Prada},
  {Pucha}, {Raichoor}, {Ram{\'\i}rez-P{\'e}rez}, {Ramirez-Solano},
  {Rashkovetskyi}, {Ravoux}, {Rocher}, {Rockosi}, {Ross}, {Rossi}, {Ruggeri},
  {Ruhlmann-Kleider}, {Sabiu}, {Said}, {Saintonge}, {Samushia}, {Sanchez},
  {Saulder}, {Schaan}, {Schlafly}, {Schlegel}, {Scholte}, {Schubnell}, {Seo},
  {Shafieloo}, {Sharples}, {Sheu}, {Silber}, {Sinigaglia}, {Siudek}, {Slepian},
  {Smith}, {Sprayberry}, {Stephey}, {Su{\'a}rez-P{\'e}rez}, {Sun}, {Tan},
  {Tarl{\'e}}, {Tojeiro}, {Ure{\~n}a-L{\'o}pez}, {Vaisakh}, {Valcin}, {Valdes},
  {Valluri}, {Vargas-Maga{\~n}a}, {Variu}, {Verde}, {Walther}, {Wang}, {Wang},
  {Weaver}, {Weaverdyck}, {Wechsler}, {White}, {Xie}, {Yang}, {Y{\`e}che},
  {Yu}, {Yuan}, {Zhang}, {Zhang}, {Zhao}, {Zheng}, {Zhou}, {Zhou}, {Zou},
  {Zou}, \& {Zu}}]{desiEDR2023}
{DESI Collaboration}, {Adame}, A.~G., {Aguilar}, J., {et~al.} 2023, arXiv
  e-prints, arXiv:2306.06308

\bibitem[{{Einasto} {et~al.}(1980){Einasto}, {Joeveer}, \& {Saar}}]{Einasto80}
{Einasto}, J., {Joeveer}, M., \& {Saar}, E. 1980, \mnras, 193, 353

\bibitem[{{Einasto} {et~al.}(2024){Einasto}, {Einasto}, {Tenjes}, {Korhonen},
  {Kipper}, {Tempel}, {Liivam{\"a}gi}, \& {Hein{\"a}m{\"a}ki}}]{EinastoM24}
{Einasto}, M., {Einasto}, J., {Tenjes}, P., {et~al.} 2024, \aap, 681, A91

\bibitem[{{Einasto} {et~al.}(2021){Einasto}, {Kipper}, {Tenjes}, {Lietzen},
  {Tempel}, {Liivam{\"a}gi}, {Einasto}, {Tamm}, {Hein{\"a}m{\"a}ki}, \&
  {Nurmi}}]{EinastoM21}
{Einasto}, M., {Kipper}, R., {Tenjes}, P., {et~al.} 2021, \aap, 649, A51

\bibitem[{{Euclid Collaboration} {et~al.}(2019){Euclid Collaboration}, {Adam},
  {Vannier}, {Maurogordato}, {Biviano}, {Adami}, {Ascaso}, {Bellagamba},
  {Benoist}, {Cappi}, {D{\'\i}az-S{\'a}nchez}, {Durret}, {Farrens}, {Gonzalez},
  {Iovino}, {Licitra}, {Maturi}, {Mei}, {Merson}, {Munari}, {Pell{\'o}},
  {Ricci}, {Rocci}, {Roncarelli}, {Sarron}, {Amoura}, {Andreon}, {Apostolakos},
  {Arnaud}, {Bardelli}, {Bartlett}, {Baugh}, {Borgani}, {Brodwin}, {Castander},
  {Castignani}, {Cucciati}, {De Lucia}, {Dubath}, {Fosalba}, {Giocoli},
  {Hoekstra}, {Mamon}, {Melin}, {Moscardini}, {Paltani}, {Radovich},
  {Sartoris}, {Schultheis}, {Sereno}, {Weller}, {Burigana}, {Carvalho},
  {Corcione}, {Kurki-Suonio}, {Lilje}, {Sirri}, {Toledo-Moreo}, \&
  {Zamorani}}]{Euclid}
{Euclid Collaboration}, {Adam}, R., {Vannier}, M., {et~al.} 2019, \aap, 627,
  A23

\bibitem[{{Fanaroff} \& {Riley}(1974)}]{Fanaroff1974}
{Fanaroff}, B.~L. \& {Riley}, J.~M. 1974, \mnras, 167, 31P

\bibitem[{{Garon} {et~al.}(2019){Garon}, {Rudnick}, {Wong}, {Jones}, {Kim},
  {Andernach}, {Shabala}, {Kapi{\'n}ska}, {Norris}, {de Gasperin}, {Tate}, \&
  {Tang}}]{Garon2019}
{Garon}, A.~F., {Rudnick}, L., {Wong}, O.~I., {et~al.} 2019, \aj, 157, 126

\bibitem[{{Horellou} {et~al.}(2018){Horellou}, {Intema}, {Smol{\v{c}}i{\'c}},
  {Nilsson}, {Karlsson}, {Krook}, {Tolliner}, {Adami}, {Benoist}, {Birkinshaw},
  {Caretta}, {Chiappetti}, {Delhaize}, {Ferrari}, {Fotopoulou}, {Guglielmo},
  {Kolokythas}, {Pacaud}, {Pierre}, {Poggianti}, {Ramos-Ceja}, {Raychaudhury},
  {R{\"o}ttgering}, \& {Vignali}}]{Horellou2018}
{Horellou}, C., {Intema}, H.~T., {Smol{\v{c}}i{\'c}}, V., {et~al.} 2018, \aap,
  620, A19

\bibitem[{{Hutschenreuter} {et~al.}(2022){Hutschenreuter}, {Anderson}, {Betti},
  {Bower}, {Brown}, {Br{\"u}ggen}, {Carretti}, {Clarke}, {Clegg}, {Costa},
  {Croft}, {Van Eck}, {Gaensler}, {de Gasperin}, {Haverkorn}, {Heald}, {Hull},
  {Inoue}, {Johnston-Hollitt}, {Kaczmarek}, {Law}, {Ma}, {MacMahon}, {Mao},
  {Riseley}, {Roy}, {Shanahan}, {Shimwell}, {Stil}, {Sobey}, {O'Sullivan},
  {Tasse}, {Vacca}, {Vernstrom}, {Williams}, {Wright}, \&
  {En{\ss}lin}}]{Hutschenreuter2022}
{Hutschenreuter}, S., {Anderson}, C.~S., {Betti}, S., {et~al.} 2022, \aap, 657,
  A43

\bibitem[{{Jackson}(1972)}]{Jackson1972}
{Jackson}, J.~C. 1972, \mnras, 156, 1P

\bibitem[{{Kim} {et~al.}(2015){Kim}, {Park}, {L'Huillier}, \& {Hong}}]{Kim15}
{Kim}, J., {Park}, C., {L'Huillier}, B., \& {Hong}, S.~E. 2015, Journal of
  Korean Astronomical Society, 48, 213

\bibitem[{{Komberg} \& {Pashchenko}(2009)}]{Komberg2009}
{Komberg}, B.~V. \& {Pashchenko}, I.~N. 2009, Astronomy Reports, 53, 1086

\bibitem[{{Konar} {et~al.}(2004){Konar}, {Saikia}, {Ishwara-Chandra}, \&
  {Kulkarni}}]{Konar2004}
{Konar}, C., {Saikia}, D.~J., {Ishwara-Chandra}, C.~H., \& {Kulkarni}, V.~K.
  2004, \mnras, 355, 845

\bibitem[{{Ku{\'z}micz} {et~al.}(2019){Ku{\'z}micz}, {Czerny}, \&
  {Wildy}}]{kuzmic19}
{Ku{\'z}micz}, A., {Czerny}, B., \& {Wildy}, C. 2019, \aap, 624, A91

\bibitem[{{Lan} \& {Prochaska}(2021)}]{Lan2021}
{Lan}, T.-W. \& {Prochaska}, J.~X. 2021, \mnras, 502, 5104

\bibitem[{{Lietzen} {et~al.}(2016){Lietzen}, {Tempel}, {Liivam{\"a}gi},
  {Montero-Dorta}, {Einasto}, {Streblyanska}, {Maraston},
  {Rubi{\~n}o-Mart{\'\i}n}, \& {Saar}}]{Lietzen16}
{Lietzen}, H., {Tempel}, E., {Liivam{\"a}gi}, L.~J., {et~al.} 2016, \aap, 588,
  L4

\bibitem[{{Liivam{\"a}gi} {et~al.}(2012){Liivam{\"a}gi}, {Tempel}, \&
  {Saar}}]{Liivamagi12}
{Liivam{\"a}gi}, L.~J., {Tempel}, E., \& {Saar}, E. 2012, \aap, 539, A80

\bibitem[{{Liu} {et~al.}(2024){Liu}, {Bulbul}, {Kluge}, {Ghirardini}, {Zhang},
  {Sanders}, {Artis}, {Bahar}, {Balzer}, {Br{\"u}ggen}, {Clerc}, {Comparat},
  {Garrel}, {Gatuzz}, {Grandis}, {Lamer}, {Merloni}, {Migkas}, {Nandra},
  {Predehl}, {Ramos-Ceja}, {Reiprich}, {Seppi}, \& {Zelmer}}]{Liu2024}
{Liu}, A., {Bulbul}, E., {Kluge}, M., {et~al.} 2024, \aap, 683, A130

\bibitem[{{Mack} {et~al.}(1998){Mack}, {Klein}, {O'Dea}, {Willis}, \&
  {Saripalli}}]{mack98}
{Mack}, K.~H., {Klein}, U., {O'Dea}, C.~P., {Willis}, A.~G., \& {Saripalli}, L.
  1998, \aap, 329, 431

\bibitem[{{Mauduit} \& {Mamon}(2007)}]{Mauduit2007}
{Mauduit}, J.~C. \& {Mamon}, G.~A. 2007, \aap, 475, 169

\bibitem[{{O'Dea} \& {Baum}(2023)}]{2023WATREV}
{O'Dea}, C.~P. \& {Baum}, S.~A. 2023, Galaxies, 11, 67

\bibitem[{{Oei} {et~al.}(2023{\natexlab{a}}){Oei}, {van Weeren}, {Gast},
  {Botteon}, {Hardcastle}, {Dabhade}, {Shimwell}, {R{\"o}ttgering}, \&
  {Drabent}}]{Oei2023}
{Oei}, M. S.~S.~L., {van Weeren}, R.~J., {Gast}, A. R.~D.~J.~G.~I.~B., {et~al.}
  2023{\natexlab{a}}, \aap, 672, A163

\bibitem[{{Oei} {et~al.}(2024){Oei}, {van Weeren}, {Hardcastle}, {Gast},
  {Leclercq}, {R{\"o}ttgering}, {Dabhade}, {Shimwell}, \& {Botteon}}]{2024Oei}
{Oei}, M. S.~S.~L., {van Weeren}, R.~J., {Hardcastle}, M.~J., {et~al.} 2024,
  arXiv e-prints, arXiv:2404.17776

\bibitem[{{Oei} {et~al.}(2023{\natexlab{b}}){Oei}, {van Weeren}, {Hardcastle},
  {Vazza}, {Shimwell}, {Leclercq}, {Br{\"u}ggen}, \&
  {R{\"o}ttgering}}]{Oeiwhim2023}
{Oei}, M. S.~S.~L., {van Weeren}, R.~J., {Hardcastle}, M.~J., {et~al.}
  2023{\natexlab{b}}, \mnras, 518, 240

\bibitem[{{O'Sullivan} {et~al.}(2018){O'Sullivan}, {Br{\"u}ggen}, {Van Eck},
  {Hardcastle}, {Haverkorn}, {Shimwell}, {Tasse}, {Vacca}, {Horellou}, \&
  {Heald}}]{OSullivan2018}
{O'Sullivan}, S.~P., {Br{\"u}ggen}, M., {Van Eck}, C.~L., {et~al.} 2018,
  Galaxies, 6, 126

\bibitem[{{O'Sullivan} {et~al.}(2019){O'Sullivan}, {Machalski}, {Van Eck},
  {Heald}, {Br{\"u}ggen}, {Fynbo}, {Heintz}, {Lara-Lopez}, {Vacca},
  {Hardcastle}, {Shimwell}, {Tasse}, {Vazza}, {Andernach}, {Birkinshaw},
  {Haverkorn}, {Horellou}, {Williams}, {Harwood}, {Brunetti}, {Anderson},
  {Mao}, {Nikiel-Wroczy{\'n}ski}, {Takahashi}, {Carretti}, {Vernstrom}, {van
  Weeren}, {Orr{\'u}}, {Morabito}, \& {Callingham}}]{Sullivan2019}
{O'Sullivan}, S.~P., {Machalski}, J., {Van Eck}, C.~L., {et~al.} 2019, \aap,
  622, A16

\bibitem[{{O'Sullivan} {et~al.}(2023){O'Sullivan}, {Shimwell}, {Hardcastle},
  {Tasse}, {Heald}, {Carretti}, {Br{\"u}ggen}, {Vacca}, {Sobey}, {Van Eck},
  {Horellou}, {Beck}, {Bilicki}, {Bourke}, {Botteon}, {Croston}, {Drabent},
  {Duncan}, {Heesen}, {Ideguchi}, {Kirwan}, {Lawlor}, {Mingo},
  {Nikiel-Wroczy{\'n}ski}, {Piotrowska}, {Scaife}, \& {van
  Weeren}}]{Sullivan2023}
{O'Sullivan}, S.~P., {Shimwell}, T.~W., {Hardcastle}, M.~J., {et~al.} 2023,
  \mnras, 519, 5723

\bibitem[{{Pillastrini}(2019)}]{Pillastrini2019}
{Pillastrini}, G. C.~B. 2019, \na, 69, 1

\bibitem[{{Planck Collaboration} {et~al.}(2020){Planck Collaboration},
  {Aghanim}, {Akrami}, {Ashdown}, {Aumont}, {Baccigalupi}, {Ballardini},
  {Banday}, {Barreiro}, {Bartolo}, {Basak}, {Battye}, {Benabed}, {Bernard},
  {Bersanelli}, {Bielewicz}, {Bock}, {Bond}, {Borrill}, {Bouchet}, {Boulanger},
  {Bucher}, {Burigana}, {Butler}, {Calabrese}, {Cardoso}, {Carron},
  {Challinor}, {Chiang}, {Chluba}, {Colombo}, {Combet}, {Contreras}, {Crill},
  {Cuttaia}, {de Bernardis}, {de Zotti}, {Delabrouille}, {Delouis}, {Di
  Valentino}, {Diego}, {Dor{\'e}}, {Douspis}, {Ducout}, {Dupac}, {Dusini},
  {Efstathiou}, {Elsner}, {En{\ss}lin}, {Eriksen}, {Fantaye}, {Farhang},
  {Fergusson}, {Fernandez-Cobos}, {Finelli}, {Forastieri}, {Frailis},
  {Fraisse}, {Franceschi}, {Frolov}, {Galeotta}, {Galli}, {Ganga},
  {G{\'e}nova-Santos}, {Gerbino}, {Ghosh}, {Gonz{\'a}lez-Nuevo}, {G{\'o}rski},
  {Gratton}, {Gruppuso}, {Gudmundsson}, {Hamann}, {Handley}, {Hansen},
  {Herranz}, {Hildebrandt}, {Hivon}, {Huang}, {Jaffe}, {Jones}, {Karakci},
  {Keih{\"a}nen}, {Keskitalo}, {Kiiveri}, {Kim}, {Kisner}, {Knox},
  {Krachmalnicoff}, {Kunz}, {Kurki-Suonio}, {Lagache}, {Lamarre}, {Lasenby},
  {Lattanzi}, {Lawrence}, {Le Jeune}, {Lemos}, {Lesgourgues}, {Levrier},
  {Lewis}, {Liguori}, {Lilje}, {Lilley}, {Lindholm}, {L{\'o}pez-Caniego},
  {Lubin}, {Ma}, {Mac{\'\i}as-P{\'e}rez}, {Maggio}, {Maino}, {Mandolesi},
  {Mangilli}, {Marcos-Caballero}, {Maris}, {Martin}, {Martinelli},
  {Mart{\'\i}nez-Gonz{\'a}lez}, {Matarrese}, {Mauri}, {McEwen}, {Meinhold},
  {Melchiorri}, {Mennella}, {Migliaccio}, {Millea}, {Mitra},
  {Miville-Desch{\^e}nes}, {Molinari}, {Montier}, {Morgante}, {Moss}, {Natoli},
  {N{\o}rgaard-Nielsen}, {Pagano}, {Paoletti}, {Partridge}, {Patanchon},
  {Peiris}, {Perrotta}, {Pettorino}, {Piacentini}, {Polastri}, {Polenta},
  {Puget}, {Rachen}, {Reinecke}, {Remazeilles}, {Renzi}, {Rocha}, {Rosset},
  {Roudier}, {Rubi{\~n}o-Mart{\'\i}n}, {Ruiz-Granados}, {Salvati}, {Sandri},
  {Savelainen}, {Scott}, {Shellard}, {Sirignano}, {Sirri}, {Spencer},
  {Sunyaev}, {Suur-Uski}, {Tauber}, {Tavagnacco}, {Tenti}, {Toffolatti},
  {Tomasi}, {Trombetti}, {Valenziano}, {Valiviita}, {Van Tent}, {Vibert},
  {Vielva}, {Villa}, {Vittorio}, {Wandelt}, {Wehus}, {White}, {White},
  {Zacchei}, \& {Zonca}}]{Planck18}
{Planck Collaboration}, {Aghanim}, N., {Akrami}, Y., {et~al.} 2020, \aap, 641,
  A6

\bibitem[{{Raychaudhury}(1989)}]{Raychaudhury89}
{Raychaudhury}, S. 1989, \nat, 342, 251

\bibitem[{{Rudnick} \& {Blundell}(2003)}]{Rudnick2003}
{Rudnick}, L. \& {Blundell}, K.~M. 2003, \apj, 588, 143

\bibitem[{{Safouris} {et~al.}(2009){Safouris}, {Subrahmanyan}, {Bicknell}, \&
  {Saripalli}}]{Safouris2009}
{Safouris}, V., {Subrahmanyan}, R., {Bicknell}, G.~V., \& {Saripalli}, L. 2009,
  \mnras, 393, 2

\bibitem[{{Sankhyayan} {et~al.}(2023){Sankhyayan}, {Bagchi}, {Tempel}, {More},
  {Einasto}, {Dabhade}, {Raychaudhury}, {Athreya}, \&
  {Hein{\"a}m{\"a}ki}}]{Sankhyayan2023}
{Sankhyayan}, S., {Bagchi}, J., {Tempel}, E., {et~al.} 2023, \apj, 958, 62

\bibitem[{{Shimwell} {et~al.}(2022){Shimwell}, {Hardcastle}, {Tasse}, {Best},
  {R{\"o}ttgering}, {Williams}, {Botteon}, {Drabent}, {Mechev}, {Shulevski},
  {van Weeren}, {Bester}, {Br{\"u}ggen}, {Brunetti}, {Callingham}, {Chy{\.z}y},
  {Conway}, {Dijkema}, {Duncan}, {de Gasperin}, {Hale}, {Haverkorn}, {Hugo},
  {Jackson}, {Mevius}, {Miley}, {Morabito}, {Morganti}, {Offringa}, {Oonk},
  {Rafferty}, {Sabater}, {Smith}, {Schwarz}, {Smirnov}, {O'Sullivan},
  {Vedantham}, {White}, {Albert}, {Alegre}, {Asabere}, {Bacon}, {Bonafede},
  {Bonnassieux}, {Brienza}, {Bilicki}, {Bonato}, {Calistro Rivera}, {Cassano},
  {Cochrane}, {Croston}, {Cuciti}, {Dallacasa}, {Danezi}, {Dettmar}, {Di
  Gennaro}, {Edler}, {En{\ss}lin}, {Emig}, {Franzen}, {Garc{\'\i}a-Vergara},
  {Grange}, {G{\"u}rkan}, {Hajduk}, {Heald}, {Heesen}, {Hoang}, {Hoeft},
  {Horellou}, {Iacobelli}, {Jamrozy}, {Jeli{\'c}}, {Kondapally}, {Kukreti},
  {Kunert-Bajraszewska}, {Magliocchetti}, {Mahatma}, {Ma{\l}ek}, {Mandal},
  {Massaro}, {Meyer-Zhao}, {Mingo}, {Mostert}, {Nair}, {Nakoneczny},
  {Nikiel-Wroczy{\'n}ski}, {Orr{\'u}}, {Pajdosz-{\'S}mierciak}, {Pasini},
  {Prandoni}, {van Piggelen}, {Rajpurohit}, {Retana-Montenegro}, {Riseley},
  {Rowlinson}, {Saxena}, {Schrijvers}, {Sweijen}, {Siewert}, {Timmerman},
  {Vaccari}, {Vink}, {West}, {Wo{\l}owska}, {Zhang}, \& {Zheng}}]{LOTSSDR2}
{Shimwell}, T.~W., {Hardcastle}, M.~J., {Tasse}, C., {et~al.} 2022, \aap, 659,
  A1

\bibitem[{{Simonte} {et~al.}(2023){Simonte}, {Andernach}, {Br{\"u}ggen},
  {Best}, \& {Osinga}}]{Simonte2023}
{Simonte}, M., {Andernach}, H., {Br{\"u}ggen}, M., {Best}, P.~N., \& {Osinga},
  E. 2023, \aap, 672, A178

\bibitem[{{Simonte} {et~al.}(2022){Simonte}, {Andernach}, {Br{\"u}ggen},
  {Schwarz}, {Prandoni}, \& {Willis}}]{Simonte2022}
{Simonte}, M., {Andernach}, H., {Br{\"u}ggen}, M., {et~al.} 2022, \mnras, 515,
  2032

\bibitem[{{Strom}(1973)}]{Strom1973}
{Strom}, R.~G. 1973, \aap, 25, 303

\bibitem[{{Vanderwoude} {et~al.}(2024){Vanderwoude}, {West}, {Gaensler},
  {Rudnick}, {Van Eck}, {Thomson}, {Andernach}, {Anderson}, {Carretti},
  {Heald}, {Leahy}, {McClure-Griffiths}, {O'Sullivan}, {Tahani}, \&
  {Willis}}]{2024possum}
{Vanderwoude}, S., {West}, J.~L., {Gaensler}, B.~M., {et~al.} 2024, \aj, 167,
  226

\bibitem[{{Venturi} {et~al.}(2022){Venturi}, {Giacintucci}, {Merluzzi},
  {Bardelli}, {Busarello}, {Dallacasa}, {Sikhosana}, {Marvil}, {Smirnov},
  {Bourdin}, {Mazzotta}, {Rossetti}, {Rudnick}, {Bernardi}, {Br{\"u}ggen},
  {Carretti}, {Cassano}, {Di Gennaro}, {Gastaldello}, {Kale}, {Knowles},
  {Koribalski}, {Heywood}, {Hopkins}, {Norris}, {Reiprich}, {Tasse},
  {Vernstrom}, {Zucca}, {Bester}, {Diego}, \& {Kanapathippillai}}]{Venturi2022}
{Venturi}, T., {Giacintucci}, S., {Merluzzi}, P., {et~al.} 2022, \aap, 660, A81

\bibitem[{{Wen} \& {Han}(2015)}]{Wen15}
{Wen}, Z.~L. \& {Han}, J.~L. 2015, \apj, 807, 178

\bibitem[{{Xu} {et~al.}(2006){Xu}, {Kronberg}, {Habib}, \& {Dufton}}]{xu2006}
{Xu}, Y., {Kronberg}, P.~P., {Habib}, S., \& {Dufton}, Q.~W. 2006, \apj, 637,
  19

\bibitem[{{York} {et~al.}(2000){York}, {Adelman}, {Anderson}, {Anderson},
  {Annis}, {Bahcall}, {Bakken}, {Barkhouser}, {Bastian}, {Berman}, {Boroski},
  {Bracker}, {Briegel}, {Briggs}, {Brinkmann}, {Brunner}, {Burles}, {Carey},
  {Carr}, {Castander}, {Chen}, {Colestock}, {Connolly}, {Crocker}, {Csabai},
  {Czarapata}, {Davis}, {Doi}, {Dombeck}, {Eisenstein}, {Ellman}, {Elms},
  {Evans}, {Fan}, {Federwitz}, {Fiscelli}, {Friedman}, {Frieman}, {Fukugita},
  {Gillespie}, {Gunn}, {Gurbani}, {de Haas}, {Haldeman}, {Harris}, {Hayes},
  {Heckman}, {Hennessy}, {Hindsley}, {Holm}, {Holmgren}, {Huang}, {Hull},
  {Husby}, {Ichikawa}, {Ichikawa}, {Ivezi{\'c}}, {Kent}, {Kim}, {Kinney},
  {Klaene}, {Kleinman}, {Kleinman}, {Knapp}, {Korienek}, {Kron}, {Kunszt},
  {Lamb}, {Lee}, {Leger}, {Limmongkol}, {Lindenmeyer}, {Long}, {Loomis},
  {Loveday}, {Lucinio}, {Lupton}, {MacKinnon}, {Mannery}, {Mantsch}, {Margon},
  {McGehee}, {McKay}, {Meiksin}, {Merelli}, {Monet}, {Munn}, {Narayanan},
  {Nash}, {Neilsen}, {Neswold}, {Newberg}, {Nichol}, {Nicinski}, {Nonino},
  {Okada}, {Okamura}, {Ostriker}, {Owen}, {Pauls}, {Peoples}, {Peterson},
  {Petravick}, {Pier}, {Pope}, {Pordes}, {Prosapio}, {Rechenmacher}, {Quinn},
  {Richards}, {Richmond}, {Rivetta}, {Rockosi}, {Ruthmansdorfer}, {Sandford},
  {Schlegel}, {Schneider}, {Sekiguchi}, {Sergey}, {Shimasaku}, {Siegmund},
  {Smee}, {Smith}, {Snedden}, {Stone}, {Stoughton}, {Strauss}, {Stubbs},
  {SubbaRao}, {Szalay}, {Szapudi}, {Szokoly}, {Thakar}, {Tremonti}, {Tucker},
  {Uomoto}, {Vanden Berk}, {Vogeley}, {Waddell}, {Wang}, {Watanabe},
  {Weinberg}, {Yanny}, {Yasuda}, \& {SDSS Collaboration}}]{York00}
{York}, D.~G., {Adelman}, J., {Anderson}, Jr., J.~E., {et~al.} 2000, \aj, 120,
  1579

\end{thebibliography}

\appendix
\section{Sky Masks}\label{sec:sky_mask}
\begin{figure}[h]
\centering
\includegraphics[width=\columnwidth]{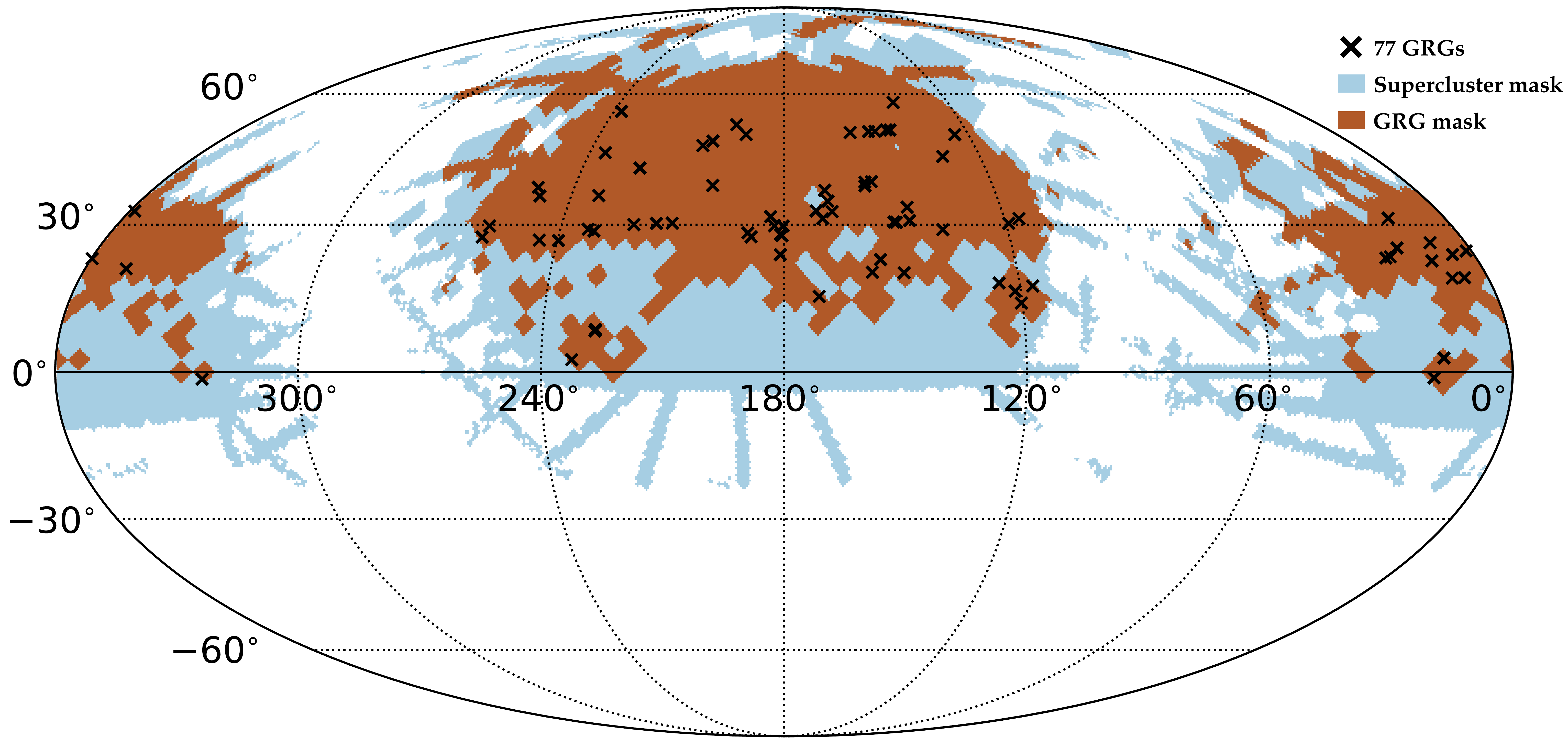}
\caption{Sky footprints of the \texttt{GRG-catalogue} and the \texttt{SCl-catalogue}. The sky coverage of the intersection of the two masks is considered for the analysis in this paper.}
\label{fig:sky_mask}
\end{figure}

\vspace{-0.2cm}
Fig.~\ref{fig:sky_mask} shows the locations of GRGs within superclusters and the sky-masks associated with the \texttt{SCl-catalogue} and \texttt{GRG-catalogue}.

\vspace{-1.0cm}
\section{Identification of giant radio galaxies in clusters}\label{sec:clust_mem}
The cluster membership of a GRG host should be treated carefully because the peculiar velocities of GRGs within a cluster can impact the identification of GRGs residing in a cluster environment. The fingers-of-god redshift distortions, induced by the peculiar velocities, may displace a member galaxy of a cluster outside the cluster's virial radius (R$_{200}$; the radius of a sphere in which the matter density is 200 times higher than the critical density of the Universe) in the redshift space. Utilizing the virial mass (M$_{200}$) of a cluster, we derived the velocity dispersion ($\sigma_{v} = \sqrt{G\mathrm{M}_{200}/3\mathrm{R}_{200}}$) induced along the line of sight. Our approach involves the initial identification of GRGs within the projected R$_{200}$ on the sky plane, followed by the identification of GRGs within the redshift range ($\bar{z} \pm \Delta z$; $\Delta z = \sigma_v (1+\bar{z})/c$) associated with cluster peculiar velocities. Here $\bar{z}$ and $c$ are the mean redshift of the cluster and the speed of light, respectively. Such GRGs are considered to be members of clusters. Among these GRGs, those lying within $3\arcsec$ of the brightest cluster galaxy (BCG) are labelled   \texttt{BCG} while the remaining ones are labelled   \texttt{Mem-Cl}. GRGs failing to meet these criteria are labelled   \texttt{No-Cl}. Corrections to the comoving distance of GRGs within a cluster to suppress the fingers-of-god redshift distortions are then done as
\begin{equation*}
\centering
\rm d_{GRG,corr} = d_{Cl} + (d_{GRG} - d_{Cl}) \frac{R_{200}}{(\sigma_{v}/H_{0})}.
\end{equation*}
Here $\rm d_{GRG,corr}$ is the corrected comoving distance to the GRG, $\rm d_{GRG}$ is the uncorrected comoving distance to the GRG, and $\rm d_{Cl}$ is the comoving distance to the cluster \citep{Liivamagi12, Bagchi17}.

\vspace{-0.3cm}
\section{Distance of giant radio galaxies from the nearest member cluster}
\begin{figure}[h]
\centering
\includegraphics[width=\columnwidth]{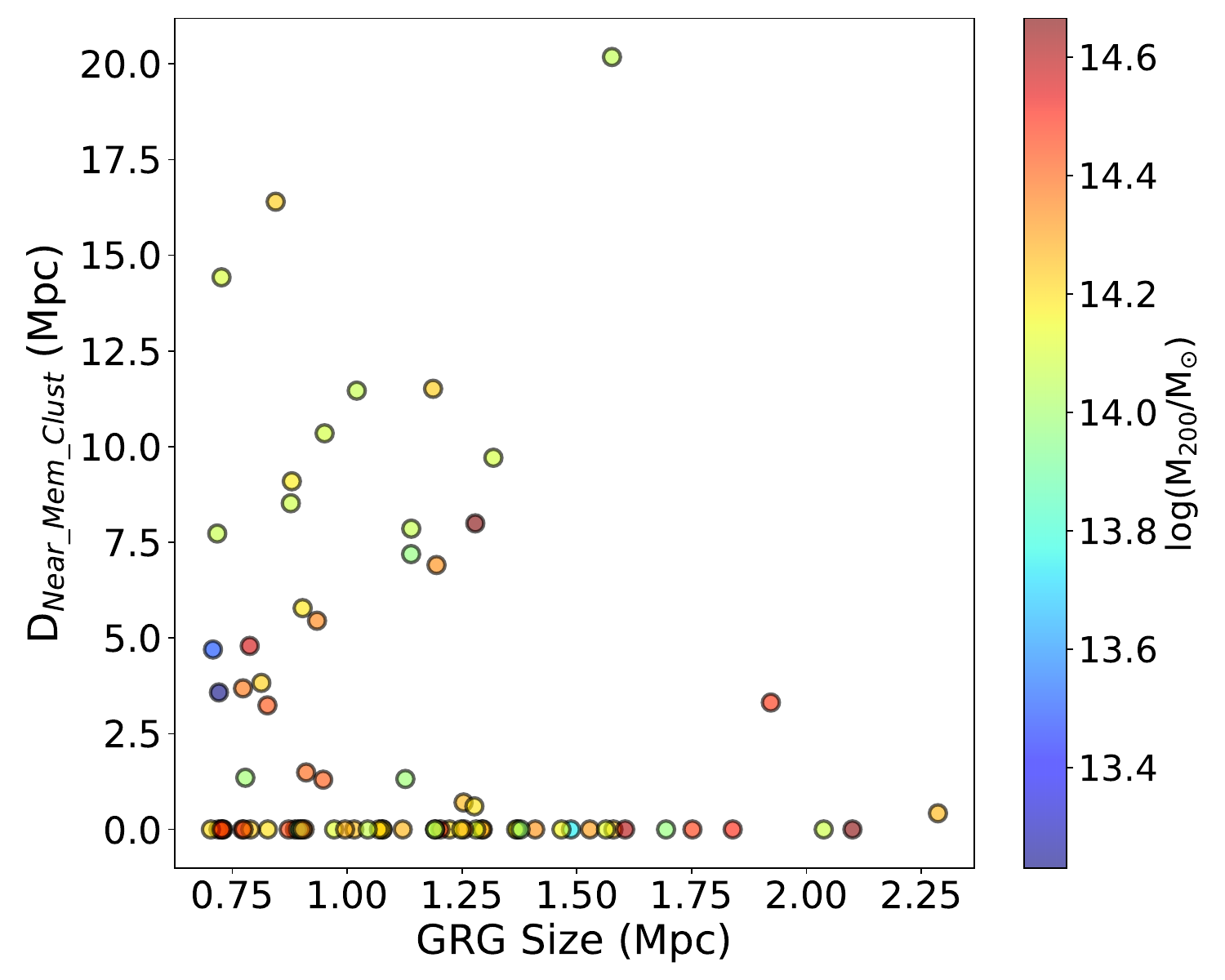}
\caption{Distance of GRGs (inside superclusters) from their nearest member cluster as a function of their size. The    galaxy cluster mass is colour-coded. Data points located in the lowest region (D$_{Near\_Mem\_Clust} = 0$) of the graph correspond to BCG-GRGs (GRGs that are BCGs).}
\label{fig:localisation}
\end{figure}

Fig.~\ref{fig:localisation} shows the distance to the nearest clusters and the sizes of GRGs within superclusters.

\vspace{-0.3cm}
\section{Magnetic field estimation} \label{sec:appendixA}
\begin{figure}
\centering
\includegraphics[width=\columnwidth]{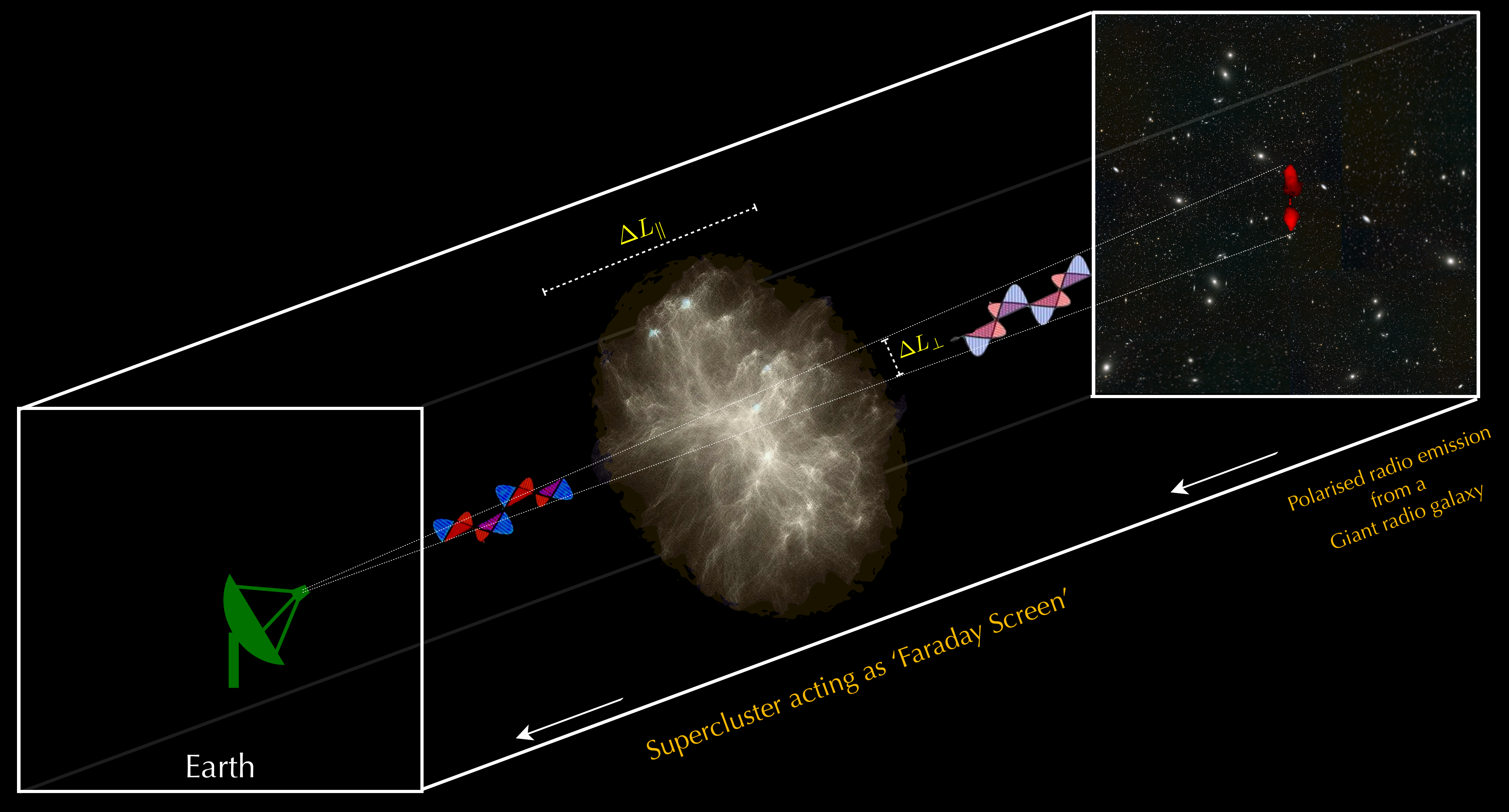}
\caption{Polarised light propagation through the supercluster's magneto-ionic medium, illustrating the resultant Faraday rotation prior to detection by terrestrial telescopes (see  Sect.~\ref{sec:grgprobe}).}
\label{fig:supeillus}
\end{figure}

Polarised electromagnetic radiation undergoes Faraday rotation when it passes through a magnetized plasma, resulting in a rotation measure (RM) determined by the   equation 
  $  \text{RM} = 0.81 \int_{0}^{L} n_e B_{\parallel} \, dl $, where 
B$_{\parallel}$ is the magnetic field component along the line of sight. The constant
0.81 is  derived from physical constants in appropriate units, where RM, n$_{e}$, B$_{\parallel}$, and $dl$ are given in   units of $\rm rad~m^{-2}$, $\rm cm^{-3}$, $\rm \muup G$, and parsec, respectively;
$\int_{0}^{L}$ represents the integral from 0 to L, where L is the total path length;
n$_{e}$ is the thermal electron number density; and $dl$ is the path length element. 
In the case of polarised light from a GRG traversing through a supercluster (in addition to the discussion in Sect.~\ref{sec:grgprobe}), we assume uniform n$_{e}$ and B$_{\parallel}$ within the supercluster and negligible magnetic field effects in a non-supercluster environment. Based on this, an estimate of magnetic field strength can be derived as B$_{\parallel} = \text{RM} / (0.81~n_e~\Delta \text{L})$. Here
$\Delta$L = $\Delta$L$_{\parallel,\mathrm{SCl}}$ and represents the path length covered by the polarised light within the supercluster (see illustration shown in Fig.~\ref{fig:supeillus}). This estimation provides insights into the magnetic field strength in superclusters under simplified conditions. These simplified conditions are applied to estimate the magnetic field inside a supercluster (also see Table~\ref{tab:magfield}). The RM values from \citet{Sullivan2023} have been corrected for the contribution from the galactic RM based on \citet{Hutschenreuter2022}, and hence they are referred to as residual rotation measure or RRM.

As shown in Table~\ref{tab:magfield}, SCl 391 and 658 each have two entries. In both cases, these measurements pertain to a single GRG, with data derived from its two lobes, whose RRMs vary notably. This variation can arise from several factors. Firstly, the local environments around each lobe may differ in magnetic field strengths and n$_{e}$, resulting in varying degrees of Faraday rotation. Secondly, the polarised light from each lobe may traverse different intervening media, such as galaxy clusters, filaments, or voids, with differing magnetic field strengths and n$_{e}$. Additionally, intrinsic differences within the lobes themselves, such as variations in magnetic field structure or plasma density, could contribute to the observed RM differences. The most likely scenario is that the polarised light from the lobes encounters significantly different intervening media, as each line of sight can intersect diverse large-scale structures (e.g. a supercluster) with varying magnetic field strength and electron densities.

\onecolumn

\section{Tables}
\begin{center}
\setlength{\tabcolsep}{4pt}
\begin{longtable}[c]{rrrrrrcrrrrrr}

\caption{Properties of 77 GRGs and their host superclusters.}

\label{tab:oeigrg} \\
\hline\hline
SCl & RA$_{\mathrm{SCl}}$ & Dec$_{\mathrm{SCl}}$ & $z_{\mathrm{SCl}}$ & CoSize$_{\mathrm{SCl}}$ &  Mass$_{\mathrm{SCl}}$ & NMem$_{\mathrm{SCl}}$ & $\delta$$_{\mathrm{SCl}}$ & RA$_{\mathrm{GRG}}$ & Dec$_{\mathrm{GRG}}$ & $z_{\mathrm{GRG}}$ &  Size$_{\mathrm{GRG}}$ & Env$_{\mathrm{GRG}}$\\
 & (deg) & (deg) & & (Mpc) & ($\rm 10^{14} M_{\odot}$) &  &  & (deg) & (deg) &  & (Mpc) & \\
 (1) & (2) & (3) & (4)& (5) & (6) & (7) & (8)  & (9) & (10) & (11)  & (12) & (13) \\
\hline
\endfirsthead

\multicolumn{13}{c}{{\tablename\ \thetable{} -- Continued from previous page}} \\
\hline\hline
SCl & RA$_{\mathrm{SCl}}$ & Dec$_{\mathrm{SCl}}$ & $z_{\mathrm{SCl}}$ & CoSize$_{\mathrm{SCl}}$ &  Mass$_{\mathrm{SCl}}$ & NMem$_{\mathrm{SCl}}$ & $\delta$$_{\mathrm{SCl}}$ & RA$_{\mathrm{GRG}}$ & Dec$_{\mathrm{GRG}}$ & $z_{\mathrm{GRG}}$ &  Size$_{\mathrm{GRG}}$ & Env$_{\mathrm{GRG}}$\\
 & (deg) & (deg) & & (Mpc) & ($10^{14} M_{\odot}$) &  &  & (deg) & (deg) &  & (Mpc) & \\
 (1) & (2) & (3) & (4)& (5) & (6) & (7) & (8)  & (9) & (10) & (11)  & (12) & (13) \\
\hline
\endhead

\hline
\multicolumn{13}{r}{{Continued on next page}} \\
\endfoot

\endlastfoot

2 & 235.1322 & 28.2855 & 0.0792 & 123.59 & 255.2 & 57 & 3.12 & \textbf{230.8622} & \textbf{28.6257} & 0.0808 & 1.08 & BCG \\
2 & 235.1322 & 28.2855 & 0.0792 & 123.59 & 255.2 & 57 & 3.12 & \textbf{239.6893} & \textbf{26.6477} & 0.0861 & 1.02 & No-Cl \\
2 & 235.1322 & 28.2855 & 0.0792 & 123.59 & 255.2 & 57 & 3.12 & \textbf{232.5428} & \textbf{29.0084} & 0.0847 & 0.72 & BCG \\
15 & 196.9778 & 52.5004 & 0.2764 & 121.59 & 151.7 & 28 & 4.36 & 195.8841 & 52.3339 & 0.2720 & 0.84 & No-Cl \\
35 & 190.7367 & 29.0072 & 0.2329 & 83.96 & 127.5 & 31 & 3.88 & 188.6948 & 27.4276 & 0.2278 & 0.97 & BCG \\
35 & 190.7367 & 29.0072 & 0.2329 & 83.96 & 127.5 & 31 & 3.88 & 189.6500 & 28.2643 & 0.2358 & 0.95 & No-Cl \\
38 & 176.4949 & 22.7656 & 0.1770 & 98.21 & 123.7 & 30 & 9.48 & 180.9321 & 23.7180 & 0.1766 & 0.73 & BCG \\
51 & 6.9235 & 19.1288 & 0.4053 & 126.32 & 111.3 & 21 & 1.57 & 5.9896 & 18.9822 & 0.4062$^\dagger$ & 1.58 & No-Cl \\
53 & 203.1943 & 48.3906 & 0.3306 & 79.58 & 110.9 & 24 & 5.35 & 202.6742 & 48.4652 & 0.3317 & 0.72 & No-Cl \\
66 & 222.5309 & 41.0626 & 0.3360 & 68.52 & 105.0 & 22 & 8.45 & 222.8934 & 42.3150 & 0.3305 & 1.30 & BCG \\
73 & 180.3112 & 26.4533 & 0.1386 & 94.41 & 101.7 & 21 & 7.23 & \textbf{180.5432} & \textbf{27.6858} & 0.1353 & 1.69 & BCG \\
73 & 180.3112 & 26.4533 & 0.1386 & 94.41 & 101.7 & 21 & 7.23 & \textbf{182.5342} & \textbf{29.7387} & 0.1430 & 0.90 & No-Cl \\
73 & 180.3112 & 26.4533 & 0.1386 & 94.41 & 101.7 & 21 & 7.23 & \textbf{180.9027} & \textbf{27.9442} & 0.1398 & 0.89 & BCG \\
74 & 227.2883 & 6.2860 & 0.0794 & 74.83 & 100.5 & 20 & 19.66 & 226.9918 & 8.2052 & 0.0775 & 0.73 & BCG \\
74 & 227.2883 & 6.2860 & 0.0794 & 74.83 & 100.5 & 20 & 19.66 & 226.8555 & 8.4957 & 0.0787 & 0.79 & No-Cl \\
81 & 145.0124 & 31.7802 & 0.3869 & 113.02 & 97.1 & 24 & 1.10 & 145.8632 & 30.8688 & 0.3999 & 1.53 & BCG \\
88 & 125.9072 & 18.2376 & 0.3009 & 106.99 & 93.7 & 21 & 7.56 & 125.1783 & 17.9105 & 0.2963 & 1.12 & BCG \\
92 & 349.1767 & 20.9235 & 0.1016 & 112.99 & 92.6 & 23 & 5.18 & 349.3324 & 20.7911 & 0.1028 & 0.73 & No-Cl \\
94 & 166.9559 & 33.3516 & 0.2499 & 70.15 & 92.4 & 20 & 11.89 & 166.7532 & 32.7736 & 0.2478 & 1.37 & BCG \\
94 & 166.9559 & 33.3516 & 0.2499 & 70.15 & 92.4 & 20 & 11.89 & 167.7293 & 35.0032 & 0.2534 & 1.22 & BCG \\
95 & 152.8222 & 31.6474 & 0.3048 & 133.11 & 92.3 & 24 & 1.51 & 150.1840 & 30.5782 & 0.3076 & 1.32 & No-Cl \\
101 & 128.9984 & 44.5972 & 0.1481 & 83.52 & 91.2 & 27 & 4.37 & 131.3629 & 44.9240 & 0.1506 & 0.81 & No-Cl \\
110 & 22.4893 & 22.3887 & 0.2338 & 73.07 & 89.5 & 23 & 4.27 & 23.5116 & 23.0341 & 0.2412$^\dagger$ & 1.37 & BCG \\
110 & 22.4893 & 22.3887 & 0.2338 & 73.07 & 89.5 & 23 & 4.27 & 22.1410 & 23.2747 & 0.2343 & 1.02 & BCG \\
139 & 323.7062 & -1.6868 & 0.3190 & 95.34 & 82.9 & 15 & 15.65 & 323.7409 & -1.4716 & 0.3192 & 1.41 & BCG \\
142 & 248.9777 & 35.5225 & 0.1630 & 63.87 & 81.8 & 21 & 7.26 & 250.3455 & 38.0360 & 0.1630 & 2.10 & BCG \\
142 & 248.9777 & 35.5225 & 0.1630 & 63.87 & 81.8 & 21 & 7.26 & 248.8438 & 36.1347 & 0.1650 & 0.93 & No-Cl \\
144 & 120.4256 & 30.2375 & 0.3203 & 78.41 & 81.5 & 16 & 5.68 & 119.1836 & 30.2619 & 0.3246 & 1.08 & BCG \\
148 & 147.2374 & 51.2899 & 0.2143 & 63.76 & 80.5 & 15 & 8.53 & 146.1975 & 51.1037 & 0.2132 & 1.20 & BCG \\
148 & 147.2374 & 51.2899 & 0.2143 & 63.76 & 80.5 & 15 & 8.53 & 145.0158 & 51.0727 & 0.2102 & 0.91 & BCG \\
154 & 231.7755 & 35.6210 & 0.2311 & 67.60 & 78.2 & 17 & 28.90 & 232.2368 & 36.2281 & 0.2263 & 1.19 & BCG \\
162 & 116.9295 & 17.0322 & 0.1861 & 51.98 & 77.4 & 16 & 14.16 & 116.7848 & 17.2907 & 0.1876 & 1.07 & BCG \\
170 & 259.8995 & 29.7333 & 0.1998 & 68.57 & 76.1 & 14 & 12.76 & 259.3549 & 29.7258 & 0.1995 & 0.88 & No-Cl \\
176 & 9.6464 & 24.5548 & 0.1490 & 75.77 & 75.4 & 21 & 4.65 & 9.3298 & 26.2201 & 0.1477 & 1.38 & BCG \\
185 & 154.9045 & 20.2379 & 0.1131 & 70.43 & 74.1 & 20 & 6.20 & 157.2733 & 20.1117 & 0.1110 & 1.25 & Mem-Cl \\
185 & 154.9045 & 20.2379 & 0.1131 & 70.43 & 74.1 & 20 & 6.20 & 154.9072 & 22.7011 & 0.1121 & 0.87 & BCG \\
198 & 17.9861 & -1.3457 & 0.1842 & 78.48 & 71.3 & 18 & 9.67 & 19.3952 & -1.1932 & 0.1855 & 1.19 & No-Cl \\
208 & 207.1101 & 48.4068 & 0.1683 & 60.80 & 70.0 & 21 & 6.54 & 205.5291 & 47.4314 & 0.1721 & 1.29 & BCG \\
239 & 156.0541 & 49.9667 & 0.1572 & 47.54 & 65.6 & 12 & 32.64 & 158.3127 & 50.4854 & 0.1610 & 0.89 & BCG \\
243 & 358.6942 & 21.2500 & 0.1702 & 66.84 & 65.0 & 16 & 7.70 & 359.7640 & 22.9057 & 0.1743 & 1.04 & BCG \\
247 & 220.0690 & 29.9014 & 0.3184 & 56.48 & 64.4 & 12 & 18.81 & 220.5093 & 30.0196 & 0.3185 & 1.58 & BCG \\
255 & 259.9947 & 29.9550 & 0.1063 & 67.98 & 63.7 & 18 & 5.11 & 260.2201 & 27.3553 & 0.1038 & 0.90 & BCG \\
270 & 15.8345 & 2.0067 & 0.1982 & 66.00 & 62.3 & 20 & 7.36 & 16.8451 & 2.7823 & 0.1957 & 1.49 & BCG \\
282 & 171.0924 & 16.2980 & 0.1722 & 98.65 & 61.6 & 17 & 4.58 & 171.0949 & 15.1661 & 0.1719 & 0.70 & BCG \\
292 & 9.2388 & 18.5378 & 0.1484 & 47.97 & 61.0 & 12 & 41.58 & 9.1443 & 18.9015 & 0.1508$^\dagger$ & 1.84 & BCG \\
324 & 115.7264 & 31.1060 & 0.3354 & 64.58 & 58.5 & 11 & 7.23 & 116.0027 & 31.2913 & 0.3408 & 1.92 & No-Cl \\
327 & 18.0800 & 25.0406 & 0.1853 & 64.14 & 58.2 & 18 & 9.36 & 18.9885 & 25.1223 & 0.1837 & 1.28 & BCG \\
355 & 168.9547 & 31.4473 & 0.2211 & 47.24 & 55.8 & 12 & 20.41 & 169.4294 & 31.2801 & 0.2206 & 1.28 & No-Cl \\
357 & 200.2504 & 38.0586 & 0.2380 & 65.29 & 55.7 & 12 & 21.91 & 200.4425 & 38.4360 & 0.2380 & 0.91 & Mem-Cl \\
359 & 150.1405 & 50.9867 & 0.3997 & 82.69 & 55.6 & 15 & 3.63 & 150.0645 & 50.7491 & 0.3990 & 0.77 & No-Cl \\
391 & 208.7158 & 27.2658 & 0.0626 & 71.41 & 53.6 & 15 & 18.87 & 214.4777 & 30.2685 & 0.0633 & 0.78 & Mem-Cl \\
400 & 190.9292 & 50.2206 & 0.3558 & 54.18 & 53.0 & 12 & 11.85 & 192.3050 & 50.0121 & 0.3559 & 1.25 & BCG \\
403 & 0.0314 & 24.6204 & 0.3865 & 107.36 & 52.8 & 14 & 7.70 & 1.3556 & 24.5018 & 0.3874 & 1.47 & BCG \\
418 & 155.3968 & 39.5241 & 0.1485 & 52.16 & 51.9 & 12 & 66.98 & 154.6892 & 39.2498 & 0.1482 & 2.04 & BCG \\
418 & 155.3968 & 39.5241 & 0.1485 & 52.16 & 51.9 & 12 & 66.98 & 156.5953 & 39.1477 & 0.1485 & 1.75 & BCG \\
422 & 158.0902 & 36.0347 & 0.1227 & 39.39 & 51.7 & 10 & 42.60 & 156.7660 & 38.3370 & 0.1243 & 1.00 & BCG \\
430 & 170.1028 & 37.6143 & 0.1048 & 57.95 & 51.0 & 19 & 13.90 & 168.3905 & 37.3744 & 0.1028 & 0.73 & BCG \\
446 & 236.1536 & 54.7598 & 0.2873 & 74.11 & 49.7 & 11 & 6.66 & 237.3650 & 55.6129 & 0.2936 & 1.28 & Mem-Cl \\
460 & 12.6146 & 22.7254 & 0.1830 & 63.18 & 48.7 & 12 & 13.59 & 11.9653 & 22.4580 & 0.1831 & 1.56 & BCG \\
462 & 125.2028 & 49.3166 & 0.1572 & 67.25 & 48.7 & 13 & 26.71 & 124.4002 & 49.9923 & 0.1591 & 1.19 & BCG \\
476 & 231.6128 & 1.8272 & 0.3286 & 79.55 & 47.3 & 10 & 4.85 & 232.4616 & 2.4209 & 0.3252 & 0.83 & BCG \\
478 & 357.2936 & 33.3015 & 0.3335 & 62.58 & 47.1 & 10 & 20.55 & 358.6730 & 32.8827 & 0.3415 & 2.29 & Mem-Cl \\
504 & 235.1780 & 46.9601 & 0.2021 & 40.50 & 45.4 & 13 & 20.80 & 235.1880 & 45.7572 & 0.2045 & 1.25 & BCG \\
506 & 15.5556 & 31.3543 & 0.2208 & 45.84 & 45.0 & 11 & 50.95 & 15.3901 & 31.3678 & 0.2198 & 1.61 & BCG \\
512 & 138.5962 & 28.9773 & 0.1807 & 64.88 & 44.4 & 19 & 6.68 & 137.3706 & 28.9551 & 0.1821 & 1.14 & No-Cl \\
525 & 178.8038 & 29.7090 & 0.1562 & 71.05 & 43.5 & 12 & 12.84 & 180.1921 & 29.7160 & 0.1606 & 0.77 & BCG \\
560 & 170.0182 & 33.0557 & 0.2079 & 66.33 & 40.9 & 14 & 7.75 & 171.1136 & 32.9353 & 0.2062 & 0.88 & No-Cl \\
582 & 149.6766 & 30.3924 & 0.3188 & 59.67 & 38.9 & 10 & 7.25 & 149.5718 & 30.4783 & 0.3223 & 1.19 & No-Cl \\
586 & 149.0897 & 20.0109 & 0.1709 & 48.20 & 38.7 & 11 & 17.12 & 149.1462 & 19.9985 & 0.1725 & 1.13 & Mem-Cl \\
604 & 181.6674 & 31.7969 & 0.2027 & 52.81 & 37.2 & 13 & 7.27 & 183.6326 & 31.6832 & 0.2020 & 0.95 & Mem-Cl \\
615 & 120.2823 & 16.1700 & 0.1050 & 61.62 & 36.2 & 14 & 7.61 & 120.2556 & 13.8312 & 0.1076 & 0.90 & BCG \\
615 & 120.2823 & 16.1700 & 0.1050 & 61.62 & 36.2 & 14 & 7.61 & 121.4295 & 16.2322 & 0.1000 & 0.79 & BCG \\
623 & 151.7800 & 50.8995 & 0.2252 & 43.27 & 35.2 & 10 & 23.39 & 152.2168 & 50.7002 & 0.2245 & 1.14 & No-Cl \\
626 & 144.9063 & 33.0645 & 0.1290 & 44.17 & 34.8 & 10 & 38.47 & 145.8875 & 33.6995 & 0.1314 & 0.83 & No-Cl \\
634 & 244.6070 & 29.7925 & 0.0960 & 70.18 & 33.3 & 11 & 13.27 & 244.7787 & 26.7573 & 0.0963 & 0.71 & No-Cl \\
658 & 5.5670 & 21.8839 & 0.0987 & 68.59 & 26.0 & 10 & 16.20 & 5.5914 & 23.7337 & 0.0979$^\dagger$ & 0.77 & BCG \\
659 & 132.7396 & 60.3174 & 0.1326 & 78.22 & 25.9 & 10 & 15.41 & 139.9519 & 57.8489 & 0.1370 & 0.72 & No-Cl \\
\hline
\end{longtable}
\tablefoot{
The subscripts `SCl' and `GRG' indicate host supercluster and GRG properties, respectively. Column 1 (SCl)  is the supercluster identification number from \citetalias{Sankhyayan2023}. The other columns include RA, Dec (right ascension and declination), $z$ (redshift), CoSize (comoving size), Size (projected physical size), Mass (supercluster mass), NMem (number of member clusters), and $\delta$ (density contrast of the supercluster). The coordinates of  GRGs  (3 in each) located in SCl~2 (see Fig. \ref{fig:conhul}) and SCl~73 are highlighted in bold.
The last column, Env$_{\mathrm{GRG}}$, categorises the environment of the GRG: BCG indicates the GRG is central within a BCG-hosted galaxy cluster (47 sources), No-Cl denotes GRGs outside any galaxy cluster (23 sources), and Mem-Cl signifies GRGs residing within a galaxy cluster (7 sources) that are not BCGs. $\rm z_{GRG}$ with $\rm \dagger$ indicates photometric redshift. Redshift information for GRGs has been sourced from their respective reporting catalogues, which are primarily derived from the SDSS \citep{York00}.
}
\end{center}

\begin{table}[ht]
  \centering
  \caption{Details of eight GRGs identified in the background of the superclusters with their respective properties.}
  \label{tab:magfield}
  \resizebox{\textwidth}{!}{
    \begin{tabular}{rrrrrrcccccrcc}
      \hline\hline
      SCl & RA$_{\mathrm{SCl}}$ & Dec$_{\mathrm{SCl}}$ & $z_{\mathrm{SCl}}$ & RA$_{\mathrm{GRG}}$ & Dec$_{\mathrm{GRG}}$ & $z_{\mathrm{GRG}}$ &Cat-ID& RRM$_{\mathrm{GRG}}$ & $\Delta$L$_{\parallel,\mathrm{SCl}}$ & $\Delta z_{\mathrm{SCl}}$ & $\Delta$L$_{\perp,\mathrm{SCl}}$ & \multicolumn{2}{c}{B$_{\parallel,\mathrm{SCl}}$(n$\rm _{e1}$, n$\rm _{e2}$)} \\
      & (deg) & (deg) & & (deg) & (deg) &  & &(rad m$^{-2}$) & (Mpc) &  & (kpc) & \multicolumn{2}{c}{($\muup$G)} \\
       (1) & (2) & (3) & (4)& (5) & (6) & (7) & (8)  & (9) & (10) & (11)  & (12)&\multicolumn{2}{c}{(13)}  \\
      \hline
      2 & 235.1322 & 28.2855 & 0.0792 & 227.24375 & 28.44117 & 0.35837 & 3015 & -0.18 & 110 & 0.0279 & 67 & 0.002 & 0.020 \\
92 & 349.1767 & 20.9235 & 0.1016 & 352.08936 & 21.76098 & 0.13186 & 9264 & 2.78 & 51 & 0.0132 & 233 & 0.068 & 0.678 \\
92 & 349.1767 & 20.9235 & 0.1016 & 352.02502 & 19.35509 & 0.18719 & 9254 & -2.74 & 51 & 0.0132 & 274 & 0.067 & 0.669 \\
176 & 9.6464 & 24.5548 & 0.1490 & 9.11250 & 25.61799 & 0.23348 & 20 & -5.27 & 61 & 0.0171 & 108 & 0.106 & 1.064 \\
265 & 204.8944 & 39.4580 & 0.1692 & 205.67650 & 38.88794 & 0.25576 & 994 & -2.89 & 40 & 0.0115 & 52 & 0.089 & 0.891 \\
265 & 204.8944 & 39.4580 & 0.1692 & 205.72708 & 37.97167 & 0.23700 & 7471 & 0.81 & 40 & 0.0115 & 54 & 0.025 & 0.252 \\
391 & 208.7158 & 27.2658 & 0.0626 & 210.18083 & 30.32194 & 0.20604 & 15905 & -0.82 & 23 & 0.0056 & 199 & 0.045 & 0.446 \\
391 & 208.7158 & 27.2658 & 0.0626 & 210.18083 & 30.32194 & 0.20604 & 7540 & 3.00 & 23 & 0.0056 & 89 & 0.164 & 1.639 \\
658 & 5.5670 & 21.8839 & 0.0987 & 4.56337 & 21.69262 & 0.30256 & 3621 & -1.32 & 32 & 0.0083 & 95 & 0.051 & 0.513 \\
658 & 5.5670 & 21.8839 & 0.0987 & 4.56337 & 21.69262 & 0.30256 & 3620 & -5.19 & 32 & 0.0083 & 104 & 0.201 & 2.014 \\
\hline

    \end{tabular}
  }
\tablefoot{
The subscripts `SCl' and `GRG' indicate foreground supercluster and background GRG properties, respectively. The columns RA, Dec, and $z$ are the right ascension, declination, and redshift, respectively. Cat-ID refers to the ID source number from the catalogue of \citet{Sullivan2023}. RRM$_{\mathrm{GRG}}$ stands for residual rotation measure (the value of rotation measure after correcting for the galactic magnetic field contribution), $\Delta$L$_{\parallel,\mathrm{SCl}}$ is the path length covered by the polarised light (from the background GRG) inside the supercluster, $\Delta$L$_{\perp,\mathrm{SCl}}$ is the approximate projected length of the polarised emission from the background GRG on the supercluster, and $\Delta$z$_{\mathrm{SCl}}$ is the redshift width of the supercluster. B$_{\parallel,\mathrm{SCl}}$ is the estimated magnetic field in the supercluster for n$\rm _{e1}$ and n$\rm _{e2}$ corresponding to the n$\rm _e$ values of 10$^{-6}$ cm$^{-3}$ and 10$^{-7}$ cm$^{-3}$, respectively (see  Sect.~\ref{sec:grgprobe} and Appendix~\ref{sec:appendixA} for further details).
}
\end{table}

\end{document}